\documentclass[preprint, 12pt, authoryear]{elsarticle}
\usepackage{graphicx} 
\usepackage{amsmath,amssymb,amsfonts}
\usepackage{algorithmic}
\usepackage{textcomp}
\usepackage{balance}
\usepackage{xspace}
\usepackage{booktabs}
\usepackage{enumitem}
\usepackage{url}
\usepackage[most]{tcolorbox}
\usepackage{subcaption}
\usepackage{multirow}
\usepackage{longtable}
\usepackage[table]{xcolor}
\usepackage[colorlinks=true,linkcolor=blue,citecolor=blue,urlcolor=blue]{hyperref}
\newcommand{\eg}{{\emph{e.g.}}\xspace}
\newcommand{\etal}{\textit{et al.}\xspace}

\def\BibTeX{{\rm B\kern-.05em{\sc i\kern-.025em b}\kern-.08em
    T\kern-.1667em\lower.7ex\hbox{E}\kern-.125emX}}

\tcbuselibrary{listings, breakable}

\usepackage{listings}
\usepackage{xcolor}

\definecolor{codebg}{RGB}{248,248,248}
\definecolor{codeframe}{RGB}{220,220,220}
\definecolor{placeholder}{RGB}{136,34,161} 
\definecolor{directive}{RGB}{0,102,204}    
\definecolor{tag}{RGB}{153,153,153}  
\lstdefinestyle{promptstyle}{
  basicstyle=\ttfamily\footnotesize,
  backgroundcolor=\color{codebg},
  frame=single,
  rulecolor=\color{codeframe},
  frameround=tttt,
  breaklines=true,
  columns=fullflexible,
  keepspaces=true,
  numbers=left,
  numberstyle=\scriptsize\color{gray},
  stepnumber=1,
  numbersep=8pt,
  showstringspaces=false,
  upquote=true,
  literate=
   *{\{}{{\textcolor{placeholder}{\{}}}1
    {\}}{{\textcolor{placeholder}{\}}}}1
    {<issue>}{{\textcolor{tag}{<issue>}}}1
    {</issue>}{{\textcolor{tag}{</issue>}}}1
    {REQUIREMENTS:}{{\textcolor{directive}{REQUIREMENTS:}}}1
    {FORMAT:}{{\textcolor{directive}{FORMAT:}}}1
}

\begin{document}
\begin{frontmatter}
\title{Quality Assurance of LLM-generated Code: Addressing Non-Functional Quality Characteristics}

\author{Xin Sun\corref{cor1}}
\ead{xin.sun@liu.se}
\author{Daniel St{\aa}hl}
\ead{daniel.stahl@liu.se}
\author{Kristian Sandahl}
\ead{kristian.sandahl@liu.se}
\author{Christoph Kessler}
\ead{christoph.kessler@liu.se}

\cortext[cor1]{Corresponding author.}
\address{Department of Computer and Information Science, Linköping University, Sweden}

\begin{abstract}
In recent years, large language models have been widely integrated into software engineering workflows, supporting tasks like code generation. While prior evaluations focus on functional correctness, there is still a limited understanding of the non-functional quality characteristics of generated code. 

Guided by the ISO/IEC 25010 quality model, this study adopts a multi-methods approach comprising three complementary elements: a literature review of 109 papers, two industry workshops with practitioners from multiple organizations, and an empirical analysis of patching real-world software issues using three LLMs. Motivated by insights from both the literature and practitioners, the empirical study examined the quality of generated patches regarding security, maintainability, and performance efficiency, which were identified as critical code-level quality attributes. 

Our results indicate that existing research primarily emphasizes security, performance efficiency, and maintainability, while other quality attributes are understudied. In contrast, practitioners prioritize maintainability and readability, warning that generated code may accelerate the accumulation of technical debt. The empirical evaluation demonstrates the instability of optimizing NFQCs through prompts in practical software engineering settings.

Overall, our findings expose a misalignment between academic focus, industry priorities, and observed model behavior, highlighting the need to integrate quality assurance mechanisms into LLM code generation pipelines to ensure that future generated code not only \emph{passes tests} but truly \emph{passes with quality.}

\end{abstract}

    
\end{frontmatter}

\section{Introduction}
\label{intro}

The prevalence of large language models for code (LLMs for code)
has brought new changes to the field of software engineering \citep{Chen2021EvaluatingLL}\citep{rozi2023codellama}. These models are tailored for understanding and processing code and can generate functionally correct outputs under appropriate prompts \citep{luo2024wizardcoder}\citep{hui2024Qwencode}\citep{jiang2023Mistral}. Currently, these powerful models have been integrated into the workflow of software engineering and are used by millions of developers worldwide. GitHub Copilot, released by GitHub and OpenAI, is an AI pair programmer that is built on Codex, which is a family of code LLMs \citep{Chen2021EvaluatingLL}\citep{github_copilot_2025}. Copilot can generate code across multiple programming languages from various types of prompts, such as natural language descriptions, function signatures, and surrounding code. In some cases, it can produce complete applications such as interactive websites or data pipelines from a single prompt. As these models continue to evolve rapidly, their potential and implications for software engineering remain in flux \citep{aidigest_timehorizons}. 

Several carefully curated metrics and benchmarks have been developed to evaluate the generation capability of code LLMs \cite{Chen2021EvaluatingLL}\cite{JimenezYWYPPN24}\cite{zhang-etal-2024-codeagent}\cite{Deng2025Nocode}\cite{Li25FEA}\cite{Du24evaluating}. \citet{Chen2021EvaluatingLL} released HumanEval, together with GitHub Copilot, to evaluate the ability of code LLMs to generate functionally correct code. SWE-bench \citep{JimenezYWYPPN24} was introduced to assess the ability of LLMs to solve real-world software engineering tasks.

While these benchmarks provide a good basis for evaluating the ability of LLMs to generate functionally correct code, they offer limited support for assessing the non-functional quality characteristics (NFQCs) of generated code.  According to the ISO/IEC 25010 quality model \citep{iso25010_2023}, software quality encompasses not only functional correctness but also NFQCs such as performance efficiency, maintainability, and security.

Due to the limited assessments of NFQCs in the current evaluation system, code LLMs may produce code with varying quality along different NFQCs,  which makes the generated code\footnote{In this study, \textbf{generated code} refers to source code generated by LLMs, including both code snippets and code patches.} unreliable and sometimes causes severe faults. For example, \citet{pearceAsleepKeyboardAssessing2022} studied the performance of GitHub Copilot in high-risk cybersecurity scenarios \cite{mitre_cwe_top25}. In their study of 1,689 programs generated across 89 scenarios, approximately 40\% were found to contain code patterns considered vulnerable, illustrating the security weakness in generated code.

In particular, it remains unclear how NFQCs of generated code are currently conceptualized and assessed in existing research, whether these studies align with practitioner concerns, and how they interact with each other in concrete code generation settings beyond functional correctness.

To address our research goal, we adopt a mixed-method design in which each component plays a distinct role, while using the ISO/IEC 25010 quality model as a common reference to structure and relate the investigation of NFQCs.

First, the literature review examines how NFQCs of generated code are currently conceptualized and evaluated in existing studies, revealing both commonly studied attributes and gaps. Secondly, the industry workshops provide an independent perspective on NFQCs by capturing how practitioners perceive, prioritize, and reason about quality concerns of generated code in real-world development settings. This step allows us to contrast research evaluation practices with practitioner expectations, rather than assuming their alignment. Finally, informed by insights from both the literature and the workshops, the empirical study focuses on a set of NFQCs that are consistently highlighted as important across research and practice. Rather than evaluating each quality characteristic in isolation, the study examines how multiple NFQCs are jointly reflected in SWE-bench under different evaluation and optimization choices. This design enables us to explore patterns for trade-offs among selected NFQCs and to illustrate how current evaluation practices may influence NFQCs when multiple quality concerns are considered simultaneously. 

Together, the mixed-methods form a complementary design that connects conceptual definitions, practitioners' perspectives, and empirical observations, providing an integrated perspective on NFQCs in generated code.

To guide our study, we defined three overarching \textbf{Research Goals (RGs)}, each addressing a different aspect. The research goals are as follows:
\begin{itemize}
    \item \textbf{RG1:} Explore how NFQCs in generated code have been addressed in existing research, and identify research gaps in this area.

    \item \textbf{RG2:} Identify the real-world expectations and pain points that practitioners have when integrating generated code in software projects.

    \item \textbf{RG3:} Characterize the observed behavior of generated code across key NFQCs identified in the previous research goals, and examine how feedback-driven strategies affect these qualities and the trade-offs among them in a real-world software engineering context.
    
\end{itemize}

For each research goal, we further formulated corresponding \textbf{Research Questions (RQs)} to guide the analysis, which are presented in the corresponding sections.

The rest of the paper is organized as follows: \autoref{back} introduces the background on code generation and related work. \autoref{literature} outlines the structure of the literature review. 
\autoref{workshop} presents the organization of the two industry workshops, detailing the participant selection, session design, and experience of practitioners. \autoref{experiment} summarizes the setup of the empirical evaluation, covering dataset selection, experiment configuration, evaluation metrics, and analysis workflow.  Additionally, we discuss the implications of our findings, highlighting challenges and opportunities for future research.
We discuss the threats to validity in \autoref{sec:threats}. Finally, \autoref{conclusion} concludes the study and outlines directions for integrating NFQCs into the quality assurance of LLM-generated code in real-world software systems.

\section{Background and Related Work}
\label{back}
This section presents the background of LLMs for code generation and related works.

\subsection{Code generation}

Code generation has become a more important application scenario since LLMs were first proposed \cite{Manna1971TowardAP}\cite{ LiventsevGHM23}. Code generation aims to automatically generate code given the problem description, high-level specifications, or existing code.
Before the advent of LLMs, code generation was limited to structured methods \cite{zef2008}. These methods usually require detailed formal specifications and several steps for generation. LLMs changed this paradigm by learning from large code corpora, thus eliminating the need for formal specifications and allowing code generation from natural language prompts \cite{luo2024wizardcoder}. 

The release of GitHub Copilot marked the beginning of integrating LLMs into daily software workflows. Since then, numerous code LLMs have emerged, demonstrating remarkable performance in generating functionally correct and meaningful code across various programming languages \cite{rozi2023codellama}\cite{hui2024Qwencode}\cite{jiang2023Mistral}\cite{Chen2021EvaluatingLL}\cite{li2022alpha}\cite{LiAZMKMMALCLZZW23}.

By 2025, many AI programming tools have emerged, and a new approach called \textit{vibe coding} has gained prominence \citep{lovable2024}\citep{bolt2024}\citep{cursor2024}\citep{windsurf2024}. In vibe coding, users do not directly engage with the code; instead, they remain entirely at the level of natural language, formulating prompts that instruct the code LLMs to generate, modify, and deploy components of the software. For example, Lovable\footnote{https://lovable.dev/} is an AI-driven platform that enables users to build full-stack applications using natural language prompts, allowing rapid prototyping and deployment without extensive coding knowledge. These advancements have significantly lowered the barrier to programming and accelerated software development processes. 

\subsection{Literature Reviews}

A review by \citet{wangAreview2023} examined LLM code generation from a different angle. By reviewing 20 existing studies, they demonstrated that while using LLMs to generate code has been widely studied, the evaluation of LLM-generated code had received relatively little attention. The study addressed functional requirements such as functional correctness, as well as NFQCs, including security, maintainability, and others. It also highlighted key limitations in current evaluation practices, such as the use of inadequate evaluation criteria, the absence of systematic and quantitative evaluation frameworks, and the lack of consideration for human involvement in the evaluation process. 

\citet{yang2024robustnesssecurityprivacyexplainability} presented a systematic literature review of 146 studies focusing on NFQCs of code LLMs. They identified and discussed six NFQCs beyond functional correctness: \textit{robustness}, \textit{security}, \textit{privacy}, \textit{explainability}, \textit{efficiency} and \textit{usability}. The study highlighted the vulnerability of LLM4Code systems to adversarial attacks, data poisoning, and privacy leaks, as well as challenges in explainability and usability. To address these issues, they proposed three complementary perspectives, which are data-centric, human-centric, and system-centric, for developing more reliable and effective LLM4Code systems in the future. This work provided a comprehensive overview for understanding the broader implications of adopting LLM4Code in software engineering. However, most of the studies included in \citet{yang2024robustnesssecurityprivacyexplainability}'s literature review focus on evaluating the models themselves, such as the architecture, training data, or prompting strategies, rather than the quality of the code generated by LLMs. In practice, a model capable of producing high-quality code in principle does not guarantee that the actual outputs will meet high standards across NFQCs. It is therefore essential to acknowledge that, in real-world practice, systematic evaluation of generated code is required to ensure the overall quality of the software system. This motivates our study, which differs from prior work by shifting the perspective from the model to its outputs, and by examining to what extent the generated code satisfies NFQCs.

\subsection{Empirical Studies}

Beyond conceptual discussions and literature reviews,  a growing body of studies has examined the NFQCs of LLM-generated code. \citet{niuEvaluatingEfficiencySource2024a} conducted a comprehensive performance study using HumanEval and MBPP benchmarks and a set of programming problems from LeetCode, an online programming practice platform. They also investigated generating efficient code using prompt engineering methods. Their findings noted that the performance can depend heavily on how a prompt is given. If not instructed, LLMs might output a straightforward solution that ensures correctness but is not efficient.

\citet{fuSecurityWeaknessesCopilotGenerated2025} found that approximately 27.3\% of generated code contained security weaknesses, spanning 43 Common Weakness Enumeration categories, including critical issues like Insufficiently Random Values and Cross-site Scripting. The study also explored the effectiveness of Copilot in fixing these vulnerabilities, demonstrating that enhanced prompts could resolve up to 55.5\% of issues.

In the area of maintainability, \citet{liuEvaluatingLanguageModels2024} applied static analysis tools such as Pylint to assess generated code and reported frequent code smells and issues, suggesting the need for human oversight.

Taken together, prior work reveals a growing effort towards understanding the NFQCs of generated code. Our study extends that effort by integrating evidence from research, practice, and empirical evaluation.

\section{Literature Review}
\label{literature}

To be able to achieve RG1 in \autoref{intro}, we first conducted a literature review to assess the current state of research and potential issues in this field. Furthermore, the literature review sought to answer the following research questions:

\begin{itemize}
    \item \textbf{RQ 3.1:} What trends can be observed in research on NFQCs of generated code?
    \item \textbf{RQ 3.2:} How have different NFQCs been examined in the existing studies, and what challenges and limitations have been identified?
\end{itemize}

\subsection{Research Design}
\label{research design}
LLMs have been utilized across various tasks. In this literature review, we focus especially on code generation. Code generation involves producing source code from natural language descriptions or limited hints, allowing NFQCs to emerge directly from the generation process. As this form of AI assistance is widely used in practice, studying it provides a representative context for the quality of generated code.

The search process of the literature review combines a structured \emph{search strategy}, clearly defined \emph{inclusion and exclusion criteria}, and the use of \emph{snowballing} to identify additional relevant literature. We included papers that examine code LLMs and AI programming tools or platforms that integrate these models, such as GitHub Copilot. To be eligible, the selected papers must explore the NFQCs of the code generated by these tools or models. To facilitate consistent discussion and comparison, based on the ISO/IEC 25010 quality model, we map the diverse quality characteristics mentioned across studies to their corresponding terms in the ISO/IEC 25010 software product quality model, thereby standardizing terminology for our analysis. The search date is January 20th, 2026.

\paragraph{Search Strategy}

We conducted the search on \textbf{Scopus}, chosen for its extensive coverage of peer-reviewed publications and inclusion of some preprints. Specifically, we limited our search to studies published between Jan. 2022 and Dec. 2025, as the majority of studies in code generation prior to 2022 were based on traditional approaches rather than LLMs, which gained major traction with the release of ChatGPT in late 2022. Our search query was formulated in two stages. First, we identified three core concepts: (1) LLMs, (2) code generation, and (3) software quality. Next, we broadened the first and third concepts using closely related terms to maximize recall while maintaining precision. For example, for \textit{LLMs}, we included terms like `AI' and `artificial intelligence'; for \textit{quality}, we included `performance' and `non-functional'. To ensure the included papers focus on code generation, we kept `code generation' unchanged.

\tcbset{colback=gray!20, colframe=black, width=\columnwidth, boxrule=0.3mm}
\begin{tcolorbox}
    \begin{itemize}
    \item LLM; large language model; AI; artificial intelligence
    \item code generation
    \item quality; performance; non-functional
\end{itemize}
\end{tcolorbox}

To ensure a high coverage of relevant papers, we deliberately avoided overly specific keywords related to individual NFQCs (\eg "robust", "efficient", "latency") in the keywords list, as these terms can vary widely across papers and might appear in different forms. Finally, we set our search query as follows:

\tcbset{colback=gray!20, colframe=black, width=\columnwidth, boxrule=0.3mm}
\begin{tcolorbox}
(``LLM'' OR ``large language models'' OR ``AI'' OR ``Artificial intelligence'' ) AND ( ``code generation'' ) AND ( ``quality'' OR ``performance'' OR ``non-functional'' )
\end{tcolorbox}

 The initial results contained many irrelevant studies, so we applied a series of filters to refine the search. We also restricted the subject area to \textit{computer science} and \textit{Engineering}, as these two categories in Scopus most accurately capture research related to LLMs and code generation. The language is restricted to English. Applying these filters, we initially retrieved 945 papers. Our inclusion and exclusion criteria are summarized in \autoref{tab:select_criteria_tab}.
 
 \begin{table}[ht]

      \caption{The selection criteria of the initial search.}
       \label{tab:select_criteria_tab}
     \centering
     \begin{tabular}{p{13cm}}
     \toprule
       \textbf{Inclusion criteria}   \\
       \hline 
       \begin{itemize}[leftmargin=*]
           \item The paper discusses one of the NFQCs or similar quality characteristics of LLM-generated code. 
           \item The paper proposes an approach, study ,or benchmark to evaluate one of the non-functional quality attributes of LLM-generated code. 
            \item The paper proposes new metrics to evaluate one of the non-functional quality attributes of LLM-generated code.
            \end{itemize}\\
             \bottomrule
        \textbf{Exclusion criteria} \\
        \toprule
        \begin{itemize}[leftmargin=*]
        \item Paper that is not written in English. 
        \item Paper that is not within the subject of Computer Science and Engineering. 
        \item Paper that appeared before 2022. 
         \end{itemize}\\
        \bottomrule

     \end{tabular}
     
 \end{table}

\begin{figure}
 \centering
 \includegraphics[width=0.5\linewidth]{ 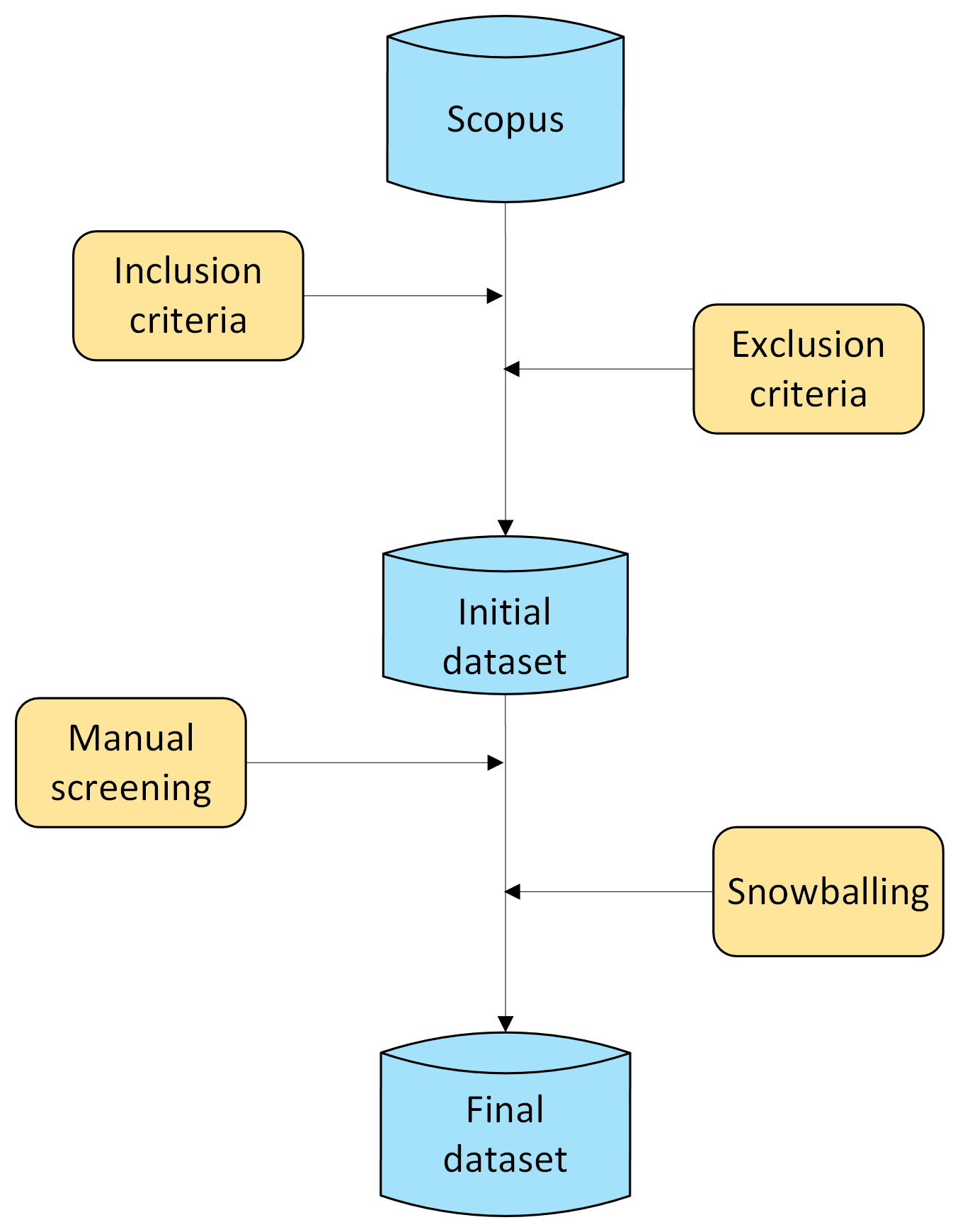}
 \caption{Search and selection process}
 \label{fig:search_process}
\end{figure}

\begin{figure}
\centering
\includegraphics[width=0.6\linewidth]{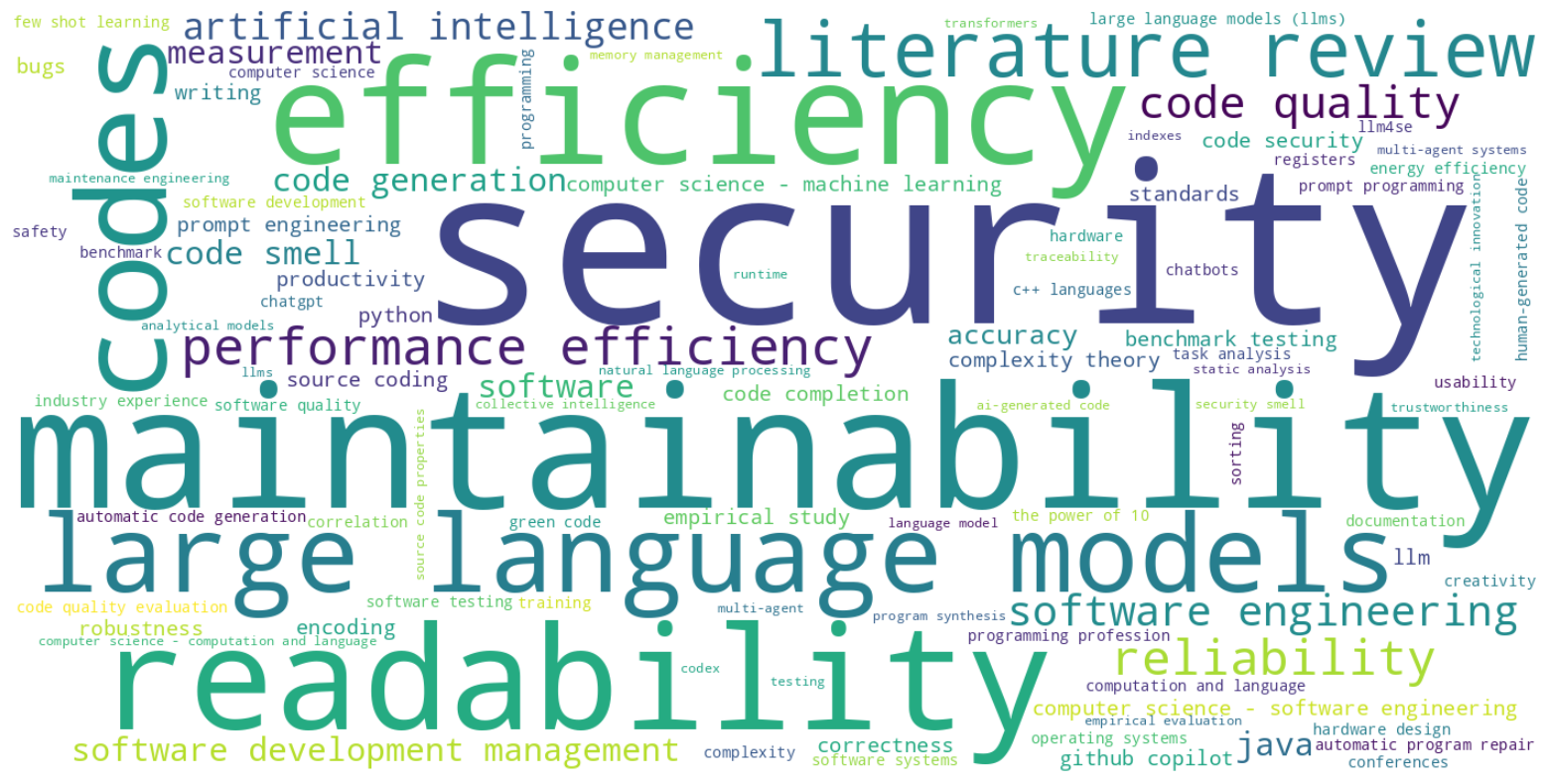}
\caption{Wordcloud of the identified literature.}
\label{fig:wordcloud}
\end{figure}

 \paragraph{Manual screen and snowballing} After the initial retrieval, we conducted a manual screening based on the title, abstract, and the keywords of each record to assess whether the study concerned generated code and discussed related to NFQCs. To ensure the quality of the included studies and our literature review, we further examined the studies along two predefined questions: \textit{Does the study address and examined quality of generated code beyond functional correctness? Does the study include empirical evaluation, systematic analysis, or concrete tooling?} Only studies that satisfied both screening questions were retained. Following this manual screening process, a total of 107 papers were included in the dataset. In addition, during the manual screening phase, we applied forward and backward snowballing to the included studies to identify further relevant literature and found two more studies. Following this process, the final dataset comprises 109 papers. The structure of the applied search and selection process is shown in \autoref{fig:search_process}. To show an overall picture of the topics covered in the final dataset, we generated a wordcloud based on the keywords of the papers, as depicted in \autoref{fig:wordcloud}

\paragraph{Data extraction and analysis}

Two authors were involved in the review and validation of the analysis. We created a spreadsheet to extract key information we wanted to analyze. In the spreadsheet, we extracted the primary information from the 109 selected papers, including title, publication year, authors, venue, evaluation metrics and tools, quality attributes addressed, LLMs evaluated, programming languages examined, and results. 

During the data extraction and analysis process, two authors mapped the quality attributes discussed in the studies to the corresponding categories defined in the ISO/IEC 25010 quality model. Studies were allowed to contribute to multiple quality attribute categories when multiple attributes were examined. The mapping results were then cross-reviewed by the two authors. Any inconsistencies were discussed together until consensus was reached. 

After completing data extraction and the initial analysis, we synthesized the collected evidence to address the research questions. For RQ 3.1, we analyzed publication years, quality attributes addressed, and programming languages examined across all included studies to identify temporal trends and shifts in research focus. For RQ 3.2, we examined the evaluation metrics and tools used, and LLMs evaluated, to characterize the current state of evaluation practices and to identify potential research gaps and opportunities.

\subsection{Literature Review Findings}
\label{liter_findings}

\paragraph{What trends can be observed?}

Following the search strategy described, 109 studies were finally included. The dataset includes 31 journal articles and 68 conference papers published across 62 distinct journals and conference venues. \autoref{tab:venue} summarizes the main publication venues and representative studies.

\begin{table}[ht]
\centering
\caption{Primary publication venues.}
\label{tab:venue}
\begin{tabular}{lp{8cm}p{3cm}}

\hline

   Venue  & Primary Publication Venue & Primary Study  \\
   \toprule
     Journal& Journal of Systems and Software & \cite{moradidakhelGitHubCopilotAI2023}\\ 
     Journal& IEEE Transactions on Software Engineering& \cite{khojahImpactPromptProgramming2025}\cite{LiuTLZZ24}\cite{zhaoSecureCodeGeneration2025}\\
     Journal& Empirical Software Engineering& \cite{borstlerDevelopersTalkingCode2023}\cite{zhengUnderstandingLargeLanguage2024}\cite{asareGitHubsCopilotBad2023} \\
     Journal&ACM Transactions on Software Engineering and Methodology &\cite{ouyangEmpiricalStudyNondeterminism2024}\cite{chenEmpiricalStudyChallenges2025a}\cite{fuSecurityWeaknessesCopilotGenerated2025} \\
     Conference &International Conference on Mining Software Repositories & \cite{moratisWriteMeThis2024a} \cite{nguyenEmpiricalEvaluationGitHub2022}\cite{siddiqQualityAssessmentChatGPT2024} \\
     Conference &International Conference on Automated Software Engineering & \cite{renMisuseMasteryEnhancing2024}\cite{siddiqSallmSecurityAssessment2024}  \\
     Conference &International Working Conference on Source Code Analysis and Manipulation &  \cite{siddiqEmpiricalStudyCode2022} \\
     Conference &International Conference on Evaluation and Assessment in Software Engineering & \cite{coignionPerformanceStudyLLMGenerated2024}\cite{dellaportaPromptPatternsAffect2025a}  \\

     Conference &ACM SIGSAC Conference on Computer and Communications Security & \cite{klemmerUsingAIAssistants2024a} \\
     Conference &International Conference on Learning Representations &\cite{ShypulaMZ0GYHNR24} \\  

\bottomrule
\end{tabular}
     \end{table}

\begin{figure}
    \centering
    \includegraphics[width=0.6\linewidth]{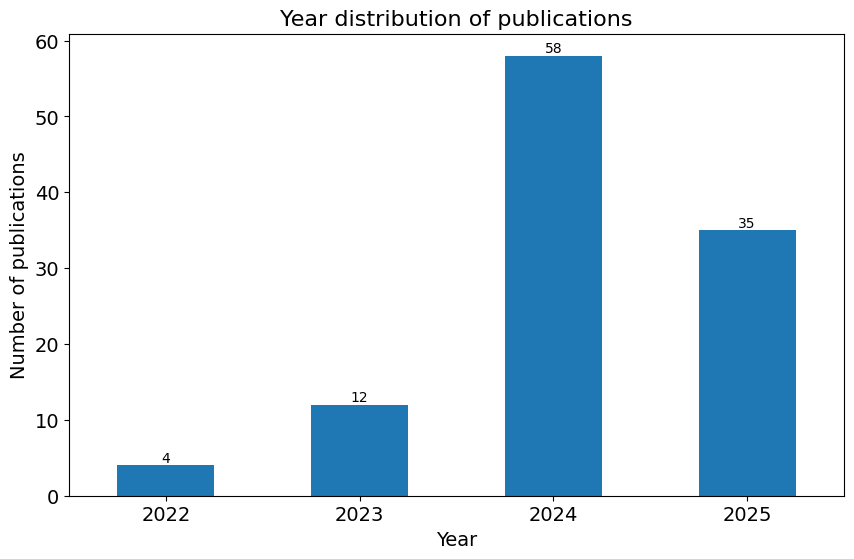}
    \caption{The year distribution of papers.}
    \label{fig:yeardistr}
\end{figure}

As shown in \autoref{fig:yeardistr}, the publication trend shows a rapid increase in studies on NFQCs of generated code between 2022 and 2024. While only four papers appeared in 2022 and 12 in 2023, the number surged to 58 in 2024. The decrease in 2025 does not necessarily indicate reduced research attention. Rather, it may reflect the fact that a number of recent studies still exist as preprints. This growth indicates that the research community has shifted from investigations of functionally correct code generation toward high-quality code generation, including several non-functional quality attributes.

\begin{figure}
    \centering
    \includegraphics[width=0.6\linewidth]{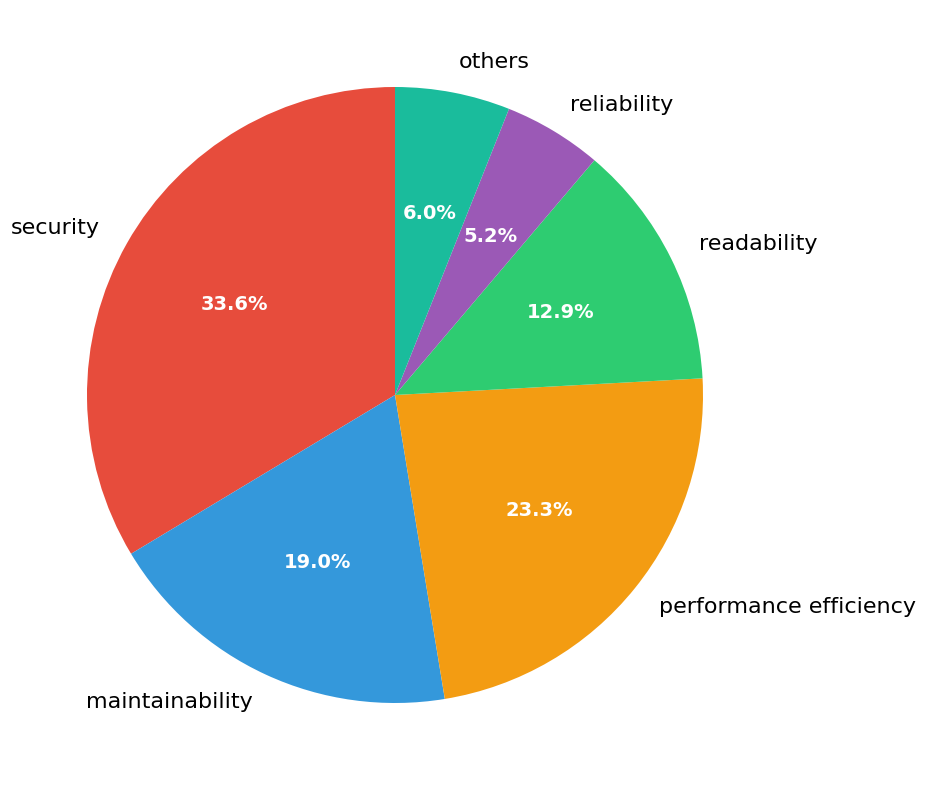}
    \caption{The distribution of papers by NFQCs.}
    \label{fig:piechart}
\end{figure}

 \autoref{fig:piechart} illustrates the distribution of NFQCs addressed by the included studies, based on our mapping to the ISO/IEC 25010 quality model. Security emerges as the most frequently examined attribute, accounting for 33.6\% of the studies. Performance efficiency (23.3\%) and maintainability (17.2\%) also receive substantial attention. In comparison, readability represents 12.9\% of the studies, while reliability (5.2\%) and modularity (1.7\%) are addressed far less often. The remaining 6.0\% of studies fall into an “others” category, covering quality aspects that are either less commonly investigated or do not map cleanly to a single ISO/IEC 25010 attribute. The distribution suggests that existing research tends to prioritize certain quality attributes over others, suggesting a potential research gap for future investigation.

In the results, we examined the distribution of programming languages evaluated across different NFQCs. Overall, Python is the most frequently studied language across all NFQCs, particularly in evaluations of security (44.1\%), performance efficiency (60.9\%), and maintainability (66.7\%). Other programming languages are less commonly examined in isolation and are more often included alongside Python within multi-language evaluation settings. These patterns suggest that current evaluation practices are influenced by language availability and execution convenience, with Python-centric benchmarks shaping much of the existing evidence on LLM code generation.

\paragraph{How NFQCs are examined and the limitations}

To answer RQ 3.2, we summarized representative studies on the evaluation of NFQCs in generated code, together with the typical evaluation methods they employ and the metrics used to reflect NFQCs, as shown in \autoref{tab:eva_studies}. 

\begin{table}[ht]
    \centering
    \caption{Thematic overview of NFQC evaluation in generated code.}
    \begin{tabular}{p{2.5cm}p{3cm}p{4cm}p{4.5cm}}
    \hline

   \textbf{NFQCs }  &\textbf{ What is evaluated}& \textbf{Typical methods}  & \textbf{Representative studies} \\
   \toprule
   Security & Presence of security vulnerabilities, security smells & Static analysis and human review& \cite{tonyLLMSecEvalDatasetNatural2023a}\cite{tonyPromptingTechniquesSecure2025}\cite{pearceAsleepKeyboardAssessing2022}\cite{siddiqEmpiricalStudyCode2022} \\

   Maintainability & code smells, readability, modularity & Static analysis, human review  & \cite{shehabEvaluatingLargeLanguage2024}\cite{moratisWriteMeThis2024a}\cite{liuRefiningChatGPTGeneratedCode2024a}\cite{leCODECHAINMODULARCODE2024b} \\

   Performance efficiency & runtime behavior, memory usage & Execution-based benchmarking (LeetCode, gem5 simulator) &\cite{shehabEvaluatingLargeLanguage2024}\cite{niuEvaluatingEfficiencySource2024a}\cite{sakibExtendingFrontierChatGPT2024a}\cite{huangEffiLearnerEnhancingEfficiency2024} \\

   Readability & readability violations, code smells & PMD, static analysis tools, human review & \cite{moratisWriteMeThis2024a}\cite{sagodiMethodologyCodeSynthesis2024a} \\

   Reliability & misuse of APIs & ROBUSTAPI benchmark & \cite{zhongCanLLMReplace2024} \\
   \bottomrule
   
    \end{tabular}

    \label{tab:eva_studies}
\end{table}

Taken together, existing studies provide a partial but structured view of how NFQCs of generated code are currently evaluated. The following paragraphs summarize the state of research across three main dimensions.

\begin{itemize}

\item \textbf{Security} A growing body of studies has observed that LLM-generated code can be functionally correct while containing security vulnerabilities \cite{PerryS0B23}\cite{KhouryABC23}\cite{pearceAsleepKeyboardAssessing2022}\cite{elgedawyOcassionally2024}\cite{Siddiq2022SecurityEvalDM}\cite{asareGitHubsCopilotBad2023}. Most existing work evaluates security using static code analysis tools, such as CodeQL, Bandit, or SonarQube, which detect security issues using the presence of predefined vulnerability patterns or insecure coding practices \cite{pearceAsleepKeyboardAssessing2022}\cite{LiuTLZZ24}. Besides the widespread use of static analysis tools, several datasets and benchmarks have been introduced to support systematic security evaluation, including SecurityEval \cite{Siddiq2022SecurityEvalDM}, LLMSevEval \cite{tonyLLMSecEvalDatasetNatural2023a} and AI Code Generators for security \cite{Natella2024AICG}. These benchmarks improve reproducible accross studies, but they primarily focus on isolated code snippets and a limited set of vulnerability types. As a result, security evaluation in the current literature remains largely constrained to what can be detected through static analysis at the code level. A small set of studies compares the security of generated code with human-written code using manual inspection \cite{asareGitHubsCopilotBad2023}. These comparisons suggest that the generated code is not uniformly worse than human-written code in terms of introducing vulnerabilities, but differences are observed in the types and severity of issues detected. 

Overall, existing studies consistently indicate security concerns in generated code, while also reflecting a strong methodological reliance on static,code-level evaluation approaches.

\item \textbf{Performance Efficiency}
Performance efficiency of generated code is mainly evaluated using runtime performance and memory usage \cite{niuEvaluatingEfficiencySource2024a}\cite{singhalNoFunEvalFunnyHow2024}\cite{coignionPerformanceStudyLLMGenerated2024}\cite{huang_effibench_2024}\cite{WaghjaleECCO24}\cite{ShypulaMZ0GYHNR24}. The studies measure these properties by executing generated solutions on standardized programming tasks, often drawn from platforms such as LeetCode. Execution time and memory consumption are then compared with human-written or reference solutions. Under this setup, many studies report that functionally correct generated code achieves performance comparable to human-written solutions. In some cases, generated solutions even outperform human-written code \cite{coignionPerformanceStudyLLMGenerated2024}. A small number of studies also consider energy consumption \cite{VartziotisDDSSH24}\cite{solovyeva2025energy}. These studies typically measure power usage during program execution and analyze it together with runtime and memory usage. The results suggest that LLMs do not optimize for energy efficiency by default. Energy improvements are usually observed only when energy consumption is explicitly included in the evaluation or optimization process.

Overall, current performance evaluations rely on a limited set of measurable indicators, mainly runtime and memory usage. These evaluations are often conducted on isolated problems. This approach supports controlled and reproducible experiments, but it also restricts performance analysis to settings where execution behavior is easy to benchmark. Interactions between performance efficiency and other quality characteristics are sometimes reported, but they are rarely examined in a systematic manner.

\item \textbf{Maintainability}
Maintainability refers to how easily code can be understood and modified by developers over time. In the literature, maintainability of generated code is most commonly evaluated using static analysis tools and predefined metrics. Tools such as SonarQube, CodeQL and PMD are frequently used to identify code smells, complexity issues, and structural problems in generated code \cite{nguyenEmpiricalEvaluationGitHub2022}\cite{Zheng2024BeyondCB}\cite{liuRefiningChatGPTGeneratedCode2024a}\cite{Yetistiren2023EvaluatingTC}. Some studies show the reports from static analysis, which report low complexity and reasonable structure in generated code \cite{nguyenEmpiricalEvaluationGitHub2022}. However, other studies combined manual review and static analysis tools \cite{liuRefiningChatGPTGeneratedCode2024a}; they report that such code can still be verbose or poorly organized, which makes it harder to maintain in the future. This suggests that static metrics reflect only part of maintainability and may overlook issues that affect long-term maintenance.

Readability is not an attribute defined ISO/IEC 25010. Existing studies evaluate readability using benchmarks or proxy metrics derived from static analysis \cite{weyssowCodeUltraFeedbackLLMasaJudgeDataset2025}. Their results suggest that code generated by GPT-4 is more readable than that generated by open-source models, likely due to the well-formatted and structured training data.  Besides readability, some studies also investigate the modularity and structure of the generated code. The study also reported that LLMs often generate code with poor modularity and structure, making it difficult for developers to understand and modify. Kang \etal~\cite{KangSK24revisiting} questioned the traditional assumption that modularity improves code quality for LLM-generated code, finding that modular code does not consistently enhance performance and may not be favored by LLMs during generation.

Several studies compare maintainability between LLM-generated and human-written code using the same analysis tools \cite{licorishComparingHumanLLM2025}\cite{Eltabakh2024QualityOA}. These comparisons yield inconsistent results. In some cases, human-written code is found to be more readable and better aligned with coding standards. In other cases, generated code appears more structured and better documented, particularly for simpler tasks. Overall, these results suggest that maintainability outcomes depend strongly on task complexity, prompt design, and evaluation method.

Taken together, current results indicate that LLMs can produce readable and partially maintainable code, but maintainability is not ensured by default. Most evaluations rely on static, code-level metrics, and human intervention remains necessary to address structural and long-term maintenance issues.

\end{itemize}

\subsection{Discussion of literature review findings}

Our analysis of the literature reveals that the research community has made some progress in evaluating the quality of generated code. However, several challenges and open gaps need to be addressed before NFQC evaluation in this context can become systematic and comparable across studies.

\begin{itemize}
    \item \textbf{Ambiguity and inconsistency in defining quality characteristics.}
Existing studies often use overlapping or loosely defined terms to describe quality characteristics, such as using readability, maintainability, or code quality interchangeably. While ISO/IEC 25010 provides a well-established taxonomy for software quality, it has not been systematically adapted to the context of generated code. As a result, studies redefine quality characteristics based on available tools or metrics, rather than on a shared conceptual framework. This lack of alignment complicates the interpretation of results and makes it difficult to determine whether different studies are evaluating comparable aspects of quality, even when similar terminology is used.

    \item \textbf{Imbalance in evaluated quality characteristics
.}
The literature mainly focuses on a small number of NFQCs, most notably performance efficiency, security, and maintainability. These characteristics are clearly important in practice, but their frequent appearance in research is also linked to the fact that they can be evaluated using existing automated, code-level tools. Other quality characteristics defined in ISO/IEC 25010 receive much less attention. This is not necessarily because they are less important, but because their evaluation depends on system context, runtime behavior, or user interaction, which are difficult to assess from isolated code artifacts. As a result, current NFQC coverage is influenced by what can be easily measured, and this may lead to a skewed view of the overall quality of generated code.

    \item \textbf{Lack of a unified quantitative evaluation framework.} 
Researchers measure the functional correctness of the generated code using accuracy or \textit{pass$@$k}, which measures how often LLMs generate a correct response on its $k_{th}$ attempt for a given problem.  In contrast, NFQC evaluation relies on a heterogeneous set of metrics, benchmarks, and tools, often tailored to individual studies. This methodological diversity limits comparability and reproducibility, as differences in reported results may stem from evaluation choices rather than underlying model behavior. More importantly, the absence of a common framework prevents joint analysis of functional correctness and multiple NFQCs, reinforcing a fragmented view of code quality that treats each characteristic in isolation.

    \item \textbf{Limited understanding of interactions and trade-offs among NFQCs.} 
Most studies assess NFQCs only for code that is already functionally correct, and some report trade-offs between correctness and individual quality characteristics. However, interactions among multiple NFQCs are rarely examined in a systematic way. Unlike traditional software engineering, where trade-offs are largely determined by design decisions and constraints, LLM-based code generation introduces new degrees of freedom through prompting strategies, model selection, and generation workflows. These mechanisms suggest that trade-offs may be influenced or steered rather than simply observed. The lack of empirical studies that explicitly investigate such controllability limits our understanding of how NFQCs jointly evolve in generated code.

\end{itemize}

In summary, current research on NFQCs in generated code is inconsistent in terminology, selective in scope, and inconsistent in evaluation methods. Addressing these gaps will require unified definitions, standardized benchmarks, and comprehensive evaluation frameworks to improve understanding of the quality of generated code.

 \section{Workshops}
 \label{workshop}

 To complement and validate the findings of our literature review, we held two interactive workshops with industry experts from several organizations. These organizations are actively exploring the integration of LLMs into large-scale software systems and seeking to ensure the reliability of LLMs' outputs. The workshops are designed to answer the following research questions:

 \begin{itemize}
     \item \textbf{RQ 4.1:} What NFQCs are considered most important by industry practitioners when evaluating generated code?

     \item \textbf{RQ 4.2:} What challenges and risks do practitioners perceive when integrating generated code into existing development workflows?
 \end{itemize}

\subsection{Workshop Design}
 \paragraph{Participants}

 We invited several groups of industry experts to participate in the workshops, with a total of 15 participants, including three of the authors. The participants are from seven different organizations in various industry areas, and most of the organizations are members of Software Center\footnote{https://www.software-center.se/} (SWC), which is an industry-academic collaborative network to accelerate industrial digitization and the adoption of novel approaches to software engineering. The participants have a wide range of professional skills, including quality-in-design, system integration and testing, CI/CD architecture, and automotive software test-driven development. Their professional experience in software engineering ranges from 3 to over 13 years, ensuring both breadth and depth of knowledge.

 \paragraph{Workshop Procedure}

 The workshops were held remotely through Microsoft Teams and divided into two sessions to accommodate different groups and foster focused discussion. In the first session, we met experts from a non-SWC organization, and later in the second session, we met SWC organization participants. During the workshop, we intentionally made the discussion open-ended, and questions were welcomed at any time to encourage spontaneous and candid input. Both sessions followed the same overall structure:

 \begin{itemize}
     \item \textbf{Presentation of our findings:} Each session began with a presentation of our literature review, highlighting the recent trends and the key findings related to NFQCs in LLM-generated code. The presentation acted as a shared foundation for discussion, providing participants with a common frame of reference for the subsequent workshop activities.

     \item \textbf{Participant sharing:} After our presentation, the participants were invited to present their ongoing projects, observed challenges, and useful practices related to this topic. This segment enabled different organizations (in the second workshop) to communicate with each other on the practical implementations and establish the synchronization between industry and academia.

     \item \textbf{Open discussion:} The discussion phase happened not only during the presentations, but also after the presentations. We did not follow a predetermined set of questions. Instead, it was facilitated in an exploratory manner, allowing participants to raise issues most relevant to their contexts and to reflect on both the presentation and their practical experience.
 \end{itemize}

 \paragraph{Data collection and analysis}
During the two sessions, two of the authors independently took notes during each session. After each workshop, the presentation slides were compiled, and all authors had meetings to consolidate the individual notes into a comprehensive workshop transcript. In cases where discrepancies arose between the two notes, these were discussed and resolved through consensus.

For data analysis, we used the definitions from ISO/IEC 25010 to systematically examine and categorize the workshop results. Unclear cases were discussed until the authors reached an agreement. Finally, the categorized data were synthesized to identify patterns and to highlight which aspects of NFQCs were emphasized during the workshops.

\subsection{Experience of Practitioners}
\label{workshop_findings}
In this section, we summarize the presentations and the discussions during the workshops. For anonymity, the companies are labeled from C1 to C7.

In the first workshop, C1 reported that in their organization, the adoption of LLM-based development tools is still at an early stage, and issues encountered during use are reported to support further model and tool improvement. Security risks in generated code were repeatedly raised, particularly for C and C++. Two QA arrangements were discussed: applying existing QA processes used for human-written code, which was seen as ensuring quality but reducing the potential efficiency improvement for using AI, and a hybrid human-AI setup, which was considered more efficient but potentially riskier if oversight is insufficient. Across discussions, maintainability emerged as the dominant concern for generated code, especially for large codebases with legacy components aged 10 to 15 years. Participants also noted that current LLMs lack sufficient contextual understanding of large codebases, limiting their effectiveness in such settings, while safety was regarded as important for them, but typically addressed at the system level rather than the code level.

In the cross-company workshop, C2, C3, and C4 presented their ongoing work relevant to NFQCs of generated code, and others joined the discussions. C2 presented several ongoing and completed projects related to AI-assisted software development, including \textit{prompt design for generating readable and clean code, generative AI for UX design, AI-based test generation and code review}. During the discussion, particular interest was expressed in the work on prompt design for generating readable and clean code, while participants also noted the persistent difficulties in measuring maintainability and raised questions about extending evaluations to larger code units.

C3 presented exploratory studies evaluating the quality of generated code using different toolchains. They shifted the focus from functional correctness toward NFQCs, especially maintainability and readability. Participants observed that small code fragments generally yield acceptable results, while larger and more complex systems exhibit poor readability and maintainability, and in some cases reproduce known vulnerabilities. They shared their mitigation strategies on this issue, which included decomposing tasks into smaller steps and adopting agent-based workflows using multiple models. The key challenges discussed were scalability, availability of high-quality training data, and prompt management. They also suggested using systematic code review, test-driven development, and learning from feedback to ensure the quality of generated code.

C4 presented work on combining formal methods with LLM-based code generation. Their approach integrates formal specifications with informal documentation as model input to achieve reliable and correct code generation. In their practices, they observed that zero-shot prompting produced unexpectedly good results. In the discussion, participants suggested that expectations for generated code may exceed those typically applied to human-written code.

During the open discussion, all participants engaged in in-depth discussions based on the work presented in the workshop. Several open questions were proposed and discussed. The discussions were summarized in \autoref{tab:summary_of_discussion}.

\begin{table}[ht]
    \centering
        \caption{Summary of key observations and open questions from the workshop discussion.}
    \small
    \begin{tabular}{p{3cm}p{5cm}p{5cm}}
    \hline
     \textbf{Topic}    & \textbf{Observation} & \textbf{Notes}  \\
     \toprule

      Functional correctness & Functional correctness was emphasized as a prerequisite for generated code. & Treated as a baseline requirement rather than a trade-off dimension. \\
      \hline
     
      Applicability of ISO/IEC 25010 NFQCs   & Not all ISO/IEC 25010 quality characteristics were considered applicable at the code level. Safety and compatibility were described as system-level properties, while maintainability and security were viewed as assessable at the code level. & Some attributes can not be evaluated at the code level. \\
      \hline 
      NFQC prioritization & Performance efficiency and security were acknowledged as important, but maintainability was consistently emphasized as the primary concern in industrial settings. & Especially relevant for large and long-term codebases. \\
      \hline 

      Maintainability metrics & Participant expressed dissatisfaction with existing maintainability metrics and questioned their ability to reflect practical maintenance effort. & Reported as a limitation of current evaluation approaches. \\ 
      \hline

     Readability & Readability was frequently raised as an important quality attribute of generated code and was commonly associated with maintainability. & Participants expressed interest in tool support for assessing readability, particularly in large projects. \\
     \hline 

     Trade-offs among NFQCs & Interactions among different NFQCs were perceived as underexplored compared to trade-offs between correctness and individual NFQCs. & Open questions raised regarding whether such trade-offs are inherent and how specific NFQCs might be prioritized or tuned during development. \\
     \bottomrule

    \end{tabular}

    \label{tab:summary_of_discussion}
\end{table}

\subsection{Discussion of Workshop findings}
\begin{itemize}
    \item \textbf{Reconsidering the scope of NFQC Evaluation.}
    The difference between code-level and system-level NFQCs has clear implications for how generated code is currently evaluated. The discussion in the workshop suggests that the frequent focus on maintainability, security, and performance efficiency in the literature is not only due to their importance, but also because these attributes are easier to measure at the code level using existing tools such as static analysis and linters. This emphasis on what is measurable risks narrowing the scope of NFQC research, as it may blur the line between what can be evaluated and what should be evaluated. In contrast, NFQCs such as usability, compatibility, and reliability depend on system context or user interaction. As a result, they are difficult to assess from isolated code snippets and are less frequently studied. Workshop discussions suggest that this focus on code-level measurability may introduce a systematic bias in current evaluations of generated code. This highlights the need for future research to more clearly justify the chosen evaluation level and to consider complementary approaches that go beyond code-level metrics.

    \item \textbf{Prioritized quality characteristics for generated code.}
    The emphasis on maintainability and readability from the practitioners highlights their concerns about the long-term sustainability of software systems that incorporate generated code. While generated code may be functionally correct, workshop discussions suggest that its increasing volume and complexity can make code comprehension more demanding, particularly when developers rely heavily on LLMs during development. This shift introduces a tension between short-term productivity gains and long-term maintenance effort. Importantly, the lack of reliable and widely accepted metrics and tools to assess the maintainability and readability in practice becomes an issue. This gap suggests that current research and tooling may be insufficient to support practitioners' needs in practice. As LLMs are increasingly used to generate larger and more complex code structures, the challenge may move from identifying defects to understanding and evolving the generated code, further underscoring the need for research that better assesses maintainability and readability in the context of generated code.

    \item \textbf{The impact of trade-offs and enhancing specific quality characteristics.}
    A significant point of discussion centered on the complex nature and implications of trade-offs among the various quality attributes within generated code. While existing research has examined trade-offs between individual NFQCs and functional correctness \cite{WaghjaleECCO24}\cite{coignionPerformanceStudyLLMGenerated2024}, implicitly assuming that other quality attributes can be considered independently or held constant. In contrast, practitioners emphasized that in real development settings, multiple NFQCs interact simultaneously, and improvements in one dimension frequently influence others in a non-linear way. This suggests that trade-offs may not be inherent properties of the code alone, but emergent outcomes shaped by development workflows, tooling choices, and usage context.

    Additionally, workshop participants did not frame trade-offs solely as constraints to be observed, but as properties that developers actively attempt to manage. Rather than asking whether NFQCs from a zero-sum relationship, discussions focused on how specific quality attributes can be prioritized under different constraints, such as emphasizing security in security-critical systems. This shift highlights a gap between current empirical evaluations, which largely report trade-offs post hoc, and industrial needs for mechanisms that enable the deliberate control and tuning of NFQCs during code generation. Addressing this gap requires moving beyond static evaluations toward experimental design and workflows that explicitly account for multi-objective optimization and quality steering in LLM-assisted development.
    
\end{itemize}

Overall, the workshop discussions reveal three main insights. First, only a subset of NFQCs can be automatically and reliably measured at the code level, as there are established tools such as static analysis and linters. Second, practitioners emphasized that maintainability and readability are important aspects given their direct impact on long-term projects and complex systems. Third, participants highlighted the need to better understand and manage trade-offs between functional correctness and individual NFQCs. Future work should explore how multiple NFQCs interact and how to manage these trade-offs effectively. Together, these insights outline the current landscape of NFQC evaluation of generated code and motivate the empirical study presented in the following section.

    Addressing the questions raised during the workshop regarding these NFQC trade-offs and interactions, the subsequent section of this paper details the establishment of our dataset and methodology specifically designed to investigate the interactions among NFQCs in generated code.

\section{Trade-offs Among NFQCs in Generated Code}
\label{experiment}

\subsection{Motivation and Research Questions}

 According to the findings from our literature review and workshops, we identified three NFQCs that consistently emerge as primary concerns in both academia and industry: maintainability, security, and performance efficiency. Thus, we deliberately scoped our experiment to these three dimensions. For performance efficiency, we further considered two measurable dimensions widely reported in the literature: runtime and memory usage. 

 Despite the prominence of these NFQCs, both our literature analysis and workshop discussions suggest that current studies largely examine these characteristics in isolation, with limited investigation into the interaction patterns and potential trade-offs among NFQCs of generated code in real-world development scenarios. This practical concern motivated us to study NFQCs with realistic software engineering tasks that reflect how generated code is actually used and evaluated. In this empirical study, we focus on analyzing the behavior and interactions of NFQCs in real-world, repository-level scenarios. This setting allows us to examine NFQCs as they arise in practice and to better understand their interaction patterns under realistic development conditions.

\begin{itemize}
    \item \textbf{RQ 5.1:} How do LLMs perform on the selected NFQCs when generating code for real-world software engineering tasks?

    \item \textbf{RQ 5.2:} Can incorporating static analysis feedback, such as results from CodeQL, into prompts improve the corresponding NFQC performance of generated code?

    \item \textbf{RQ 5.3:} Does improving one NFQC result in trade-offs with other NFQCs?
\end{itemize}

\subsection{Experiment Design}

\begin{figure*}[ht]
    \centering
    \includegraphics[width=1\linewidth]{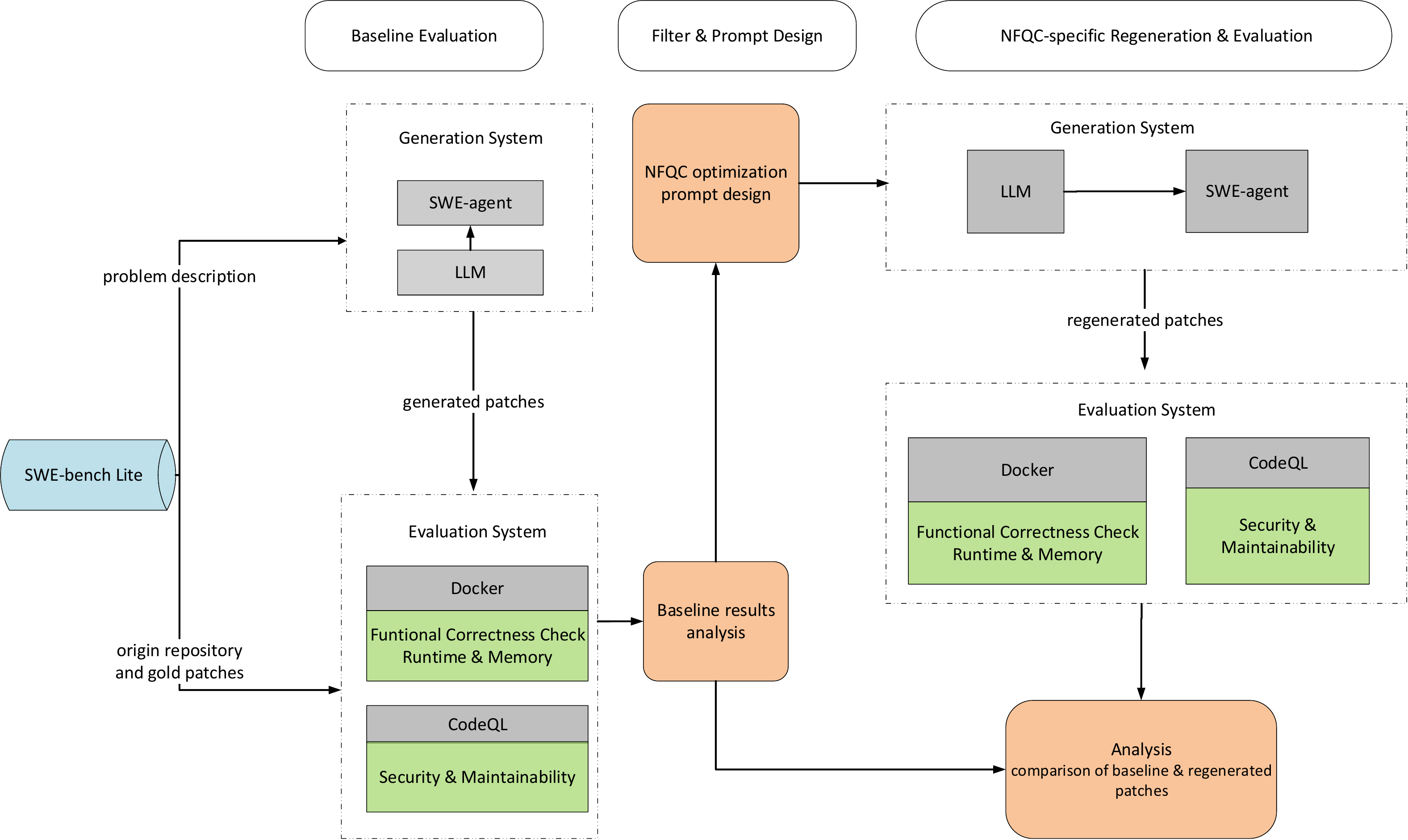}
    \caption{Overview of the experiment procedure. The workflow is divided into three main stages: (1) \textbf{Baseline Evaluation}, where initial patches were generated using SWE-agent and different LLMs, generated and benchmark gold patches are evaluated with Docker and CodeQL; (2) \textbf{Filter and Prompt Design}, where baseline results are analyzed and NFQC-specific prompts are constructed; and (3) \textbf{NFQC-specific Regeneration and Evaluation}, where patches are regenerated using NFQC-specific prompts, re-evaluated and compared against the baseline to analyze improvements and potential trade-offs.}
    \label{fig:workflow}
\end{figure*}

To address these research questions, we designed a three-stage experimental pipeline, as summarized in \autoref{fig:workflow}.
In the \textbf{first stage (Baseline Evaluation)}, we generated patches for a benchmark dataset of real-world software issues \cite{JimenezYWYPPN24} and evaluated both the generated and the benchmark gold patches. This stage established baseline measurements of functional correctness, runtime, memory usage, and static analysis results related to security and maintainability. In the \textbf{second stage (Filter and Prompt Design)}, we identified a set of comparable, correctly resolved instances across models and used the corresponding baseline results to design targeted prompts for NFQC-specific optimization. In the \textbf{final stage (NFQC-specific Regeneration and Evaluation)}, we regenerated the patches using the new prompts, evaluated them under the same conditions, and compared the results against the baseline to analyze improvements and trade-offs among different NFQCs.

\paragraph{Dataset Selection}

The goal of the experiment is to evaluate the NFQCs of generated code and explore how different NFQCs interact in functionally correct, repository-level code changes produced by LLMs. To this end, we selected \emph{SWE-bench Lite} \cite{JimenezYWYPPN24}, which is widely used by the community, as the evaluation benchmark for our experiment. Each task in SWE-bench Lite is derived from a real GitHub issue and requires generating a patch that passes the corresponding test suite. The resulting code changes typically span multiple files and interact with existing project dependencies, which makes this benchmark particularly suitable for assessing NFQCs at the repository level. In our experiments, we used the \emph{test} set of SWE-bench Lite, comprising 300 issue-pull request pairs from 11 Python repositories, predominantly drawn from Django (127 instances) and SymPy (80 instances). Each instance is evaluated by executing unit tests, using the post-pull request behavior as the reference. 

\begin{figure}
    \centering
    \includegraphics[width=0.98\linewidth]{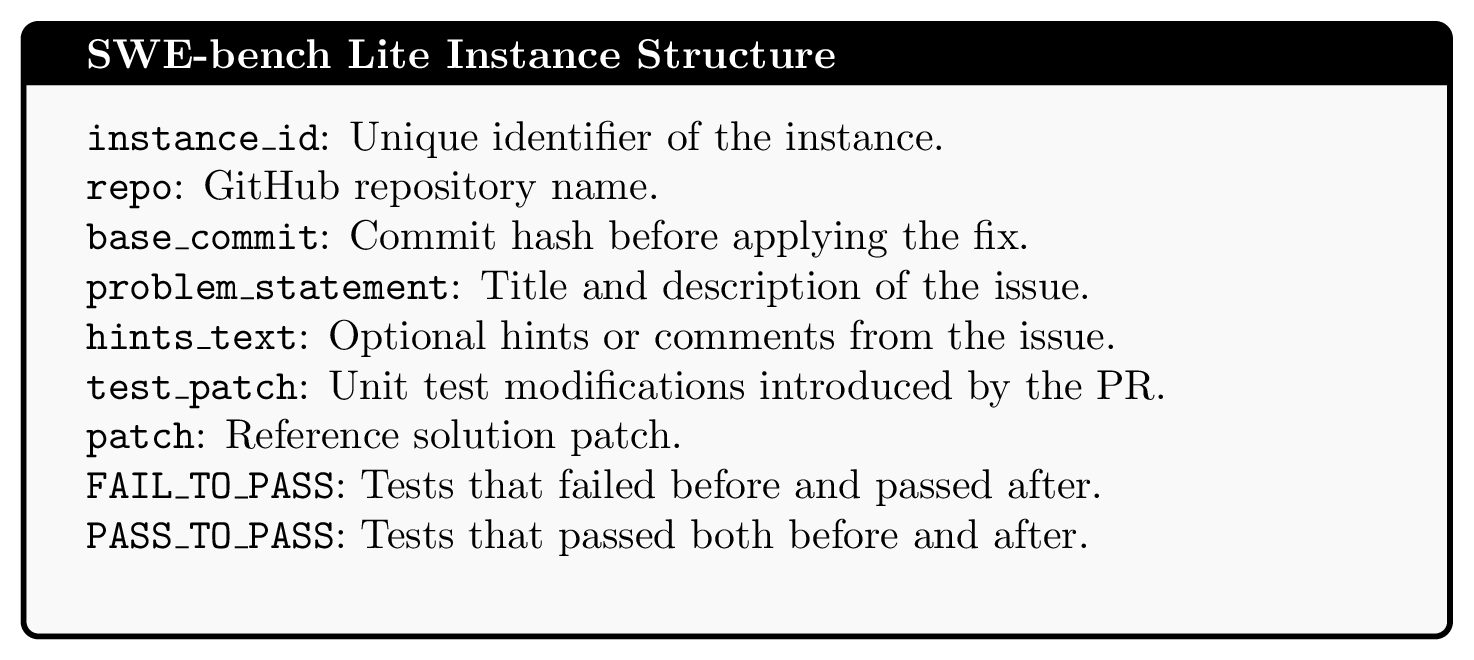}
    \caption{SWE-bench Lite Instance Structure}
    \label{fig:SWE-lite_instance}
\end{figure}

In addition, SWE-bench provides an established evaluation infrastructure, including the SWE-agent framework, which supports reproducible and automated patch generation and verification. This allows us to conduct our experiments within a standardized setup while focusing our analysis on the NFQCs of generated patches. \autoref{fig:SWE-lite_instance} shows the structure of each instance in SWE-bench Lite.

\paragraph{Model Selection}

While LLMs deliver promising performance in generating basic functions and algorithms, their performance on the SWE-bench family benchmarks remains limited. As of August 20, 2025, Claude 4 Opus achieves a resolved rate of 67.60\% on SWE-bench, and Claude 4 Sonnet reports 66.93\% (submission on May 21, 2025) and 56.67\% (submission on May 26, 2025) on SWE-bench and SWE-bench Lite, respectively.\footnote{https://github.com/SWE-bench/experiments} 

For model selection, we reviewed all the submissions on the SWE-bench Lite leaderboard and aimed to minimize confounding factors unrelated to the models themselves. To this end, we decided to use the SWE-agent framework as the generation pipeline. SWE-agent is an agent-based system designed to automate the generation process in LLM code generation. It enables LLMs to use tools, interact with the environment, and produce patches step by step, without human instruction. The use of SWE-agent also ensures that the generation process is fully driven by the LLM, thereby reducing potential biases introduced by manual intervention. This design makes comparisons across models more fair and enhances the reliability of the evaluation. To examine how models differ in their NFQCs improvement potential, we selected three models that combine representativeness and contrastiveness:

\begin{itemize}
    \item Claude 4 Sonnet: A mid-size model in the Claude 4 family. It demonstrates strong coding and reasoning abilities with enhanced instruction following and extended context support.

    \item DeepSeek-Reasoner: A reasoning-augmented model that generates chain-of-thoughts to improve the generation accuracy. It shows competitive performance on reasoning and code-related tasks compared to state-of-the-art closed-source models.

    \item GPT-4o: OpenAI's LLM, released in 2024. It achieves performance on par with GPT-4 Turbo in code generation, while offering faster inference and lower API costs. 
\end{itemize}
These models cover closed-source and open-source models, providing a more representative and generalizable comparison for evaluating NFQC in generated code.

\paragraph{Generation Strategy}

\begin{figure}
    \centering
    \includegraphics[width=0.98\linewidth]{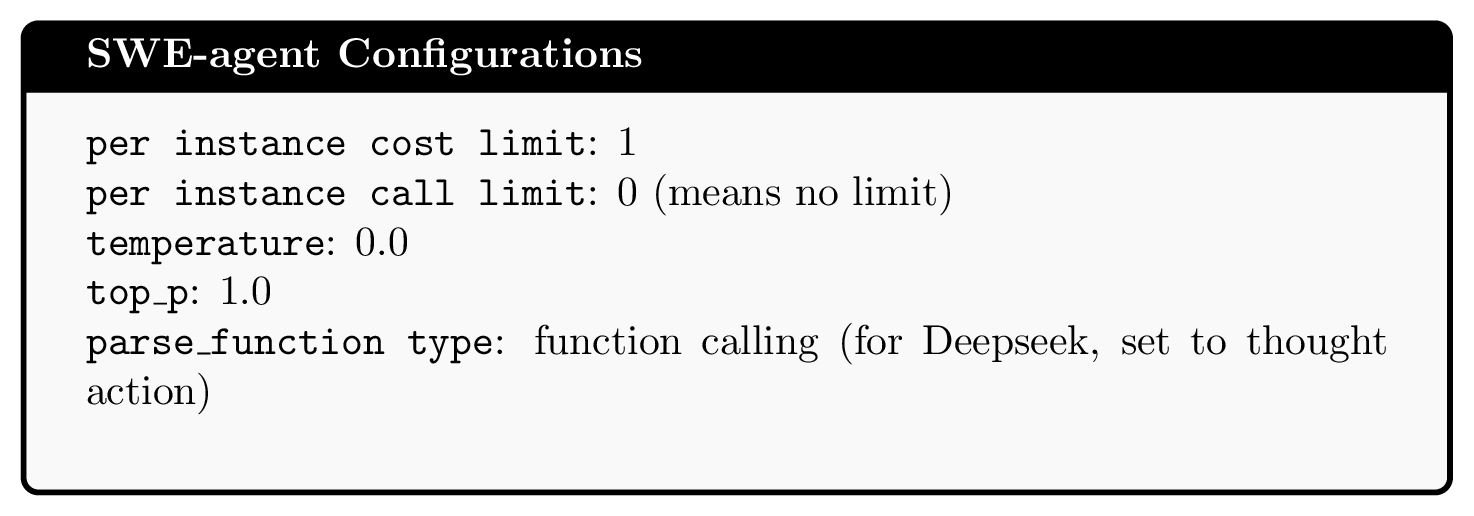}
    \caption{SWE-agent Configurations}
    \label{fig:configurations}
\end{figure}

In our experiment, we employed SWE-agent \cite{yang2024sweagent} as the generation framework. Each model has its own identical configuration parameters, and interacts with the environment through command-line tool. The agent supports various stopping criteria and cost control strategies. In the first stage, we set a cost limit of \$1 per instance for each model, without restricting the number of retries. This configuration balanced computational cost with the ability to maximize resolved instances, which provides more baseline resolved patches for subsequent analysis. In the final stage, we limited the number of retries to 1, where the agent produced one patch per instance before moving to the next instance. This multi-round generation design more closely mirrors how humans interact with LLMs in practice \cite{Zhong0S24}. The configuration is shown in \autoref{fig:configurations}.

\paragraph{Evaluation System}
The evaluation system was designed to assess the generated patches across three NFQC dimensions: \textbf{performance efficiency} (including runtime and memory usage), \textbf{security}, and \textbf{maintainability}, along with functional correctness as a prerequisite. Thus, we measured four metrics in the experiment. It consisted of two main components: The first component focused on functional verification and performance measurement. For each instance, we used Docker to create an isolated environment, cloned the corresponding repository, applied the generated patch, and executed the test cases provided by SWE-bench Lite. This step verified whether the patch was functionally correct and resolved the given issue. For patches that resolved the given issues, we further measured the total execution time required to run the entire test suite and the peak memory usage during the execution. 

The second component performed static code analysis to evaluate security and maintainability. We selected CodeQL due to its widespread industrial usage and ability to yield reproducible results. Specifically, we performed holistic repository analysis rather than incremental scans to ensure comprehensive detection. We applied the official Python security and quality query suites released by GitHub for static analysis of source code. Following the official CodeQL documentation, we filtered these suites based on their metadata tags (e.g., \textit{maintainability}, \textit{security}). This screening resulted in a specialized evaluation set comprising 39 maintainability queries and 46 security queries. \autoref{tab:maintainability_rules} lists the most frequently triggered maintainability rules, providing a granular view of the technical debt and maintainability risks identified during our experiments. We report the number of issues triggered by the patches, categorized by CodeQL's severity levels: \textit{Error} (critical flaws), \textit{Warning} (problematic patterns), and \textit{Recommendation} (suggested improvements). Additionally, to mitigate potential tool-bias and strengthen our findings, we also reported \textbf{patch size (Lines of Code, LOC)}.

\paragraph{Experimental Setup}

Our experiment was conducted on a laptop with an AMD Ryzen 7 CPU and 16GB RAM. The operating system was Ubuntu 22.04 with Python 3.11. We used SWE-agent v1.1.0, the latest release available as of May 2025, along with SWE-bench. 

The details of the three stages are as follows:
\begin{itemize}
    \item \textbf{Baseline evaluation:} We first established baseline measurements for the original repositories and SWE-bench Lite gold patches across the selected NFQCs using the evaluation system described earlier.  These measurements established the baseline results for subsequent comparison. After that, we generated the initial patches using the SWE-bench Lite and the SWE-agent. SWE-agent functioned as a unified framework that coordinated the patch generation process across different LLMs. Since the official SWE-bench Lite leaderboard already provides submissions using SWE-agent with Claude 4 Sonnet and GPT-4o, we reused the same configurations to reproduce the results. For DeepSeek-Reasoner, although no official submission was available, we used the same configuration to maintain fairness across models. Thus, in this stage, the configurations were identical across experiments, with the only difference being the LLM used. 

    After patch generation, the SWE-agent output the generated patches corresponding to the instances in the dataset. These patches were then evaluated through the evaluation system, performing the functional correctness check and the NFQC evaluation. An instance was considered \textit{resolved} if its generated patch successfully fixed the issue described in the prompt. For all resolved instances, we additionally measured the runtime and peak memory usage during the functional correctness check. The output from this stage included both the generated patches and the evaluation results of the original repositories with and without the generated patches applied. 

    \item \textbf{Filter and prompt design:} Next, we identified the resolved instances for each model and derived their intersection, meaning we retained only those instances that were successfully resolved by all three models. This ensured that subsequent analyses compared the same issues under identical functional conditions. For these common instances, we analyzed the corresponding NFQC evaluation results from CodeQL and recorded the number and type of newly introduced maintainability and security issues. Using these results, we designed NFQC optimization prompts by adapting prompts from prior studies and SWE-bench to approximate how developers might provide targeted feedback to an LLM in practice \cite{yang2024sweagent}. Building on the prompts used for the initial generation, the optimized prompts introduced two additional blocks: \textit{prev\_patch\_block} and \textit{feedback\_block}. The \textit{prev\_patch\_block} provided the correct patch generated in the first round, while the \textit{feedback\_block} gave the feedback on the specific quality dimension. For security and maintainability, the feedback block contained information derived from CodeQL reports to help the model locate the issue, such as the triggered ruleIds, a short description of the rule, and their corresponding locations, together with the targeted NFQC to be improved. For runtime and memory optimization, the feedback block only specified the dimension to be improved, without providing more detailed guidance.
    The structure of the NFQC optimization prompts is shown in \autoref{lst:NFQC_prompt}.

    \begin{lstlisting}[style=promptstyle, caption={NFQC-specific optimization prompt}, label={lst:NFQC_prompt}]
PROMPT_TMPL_BASE = """Your previous patch is functionally correct. You must re-implement its behavior directly on the base commit, while focusing on improving **{focus}**. Always generate a unified diff against the ORIGINAL BASE COMMIT.

<issue>
{problem_statement}
</issue>

{prev_patch_block}
{feedback_block}
REQUIREMENTS:
- Output **ONLY a unified diff** (git patch) with file paths relative to the repo root.
- Do NOT include explanations, markdown fences, numbering, or shell commands.
- Keep changes minimal; do NOT modify tests.
{requirement_line}
- If a trade-off arises, prioritize functional correctness.
- Re-implement the behavior of the previous patch on the base commit, preserving functionality.

FORMAT:
- Start with lines like:  diff --git a/<path> b/<path>
- Use standard unified diff hunks: @@ -old,+new @@
- Ensure the patch applies cleanly.

Now produce ONLY the patch:
"""
\end{lstlisting}

\item \textbf{NFQC-specific regeneration and evaluation:} In this stage, for each model and each instance in the intersection set, we applied the designed optimization prompts separately, producing four regenerated patches per instance per model. Every regenerated patch was evaluated with the same pipeline as in the first stage: functional correctness in isolation, followed by measurement of total runtime and peak memory if resolved, and a holistic CodeQL scan to examine security and maintainability. We then compared these outcomes against the corresponding unoptimized results for the same instance and model to quantify the improvements and assess trade-offs across NFQCs. (As in the baseline, only patches that resolve the original issue were included in comparisons.)

\end{itemize}

\subsection{Experimental Results}
\label{experiment_results}

This section presents the results of our empirical study, following the research questions and the stages of our experimental design.

\paragraph{Functional correctness of the generated patches}

\begin{figure}[hb]
    \centering
    \includegraphics[width=0.7\linewidth]{ 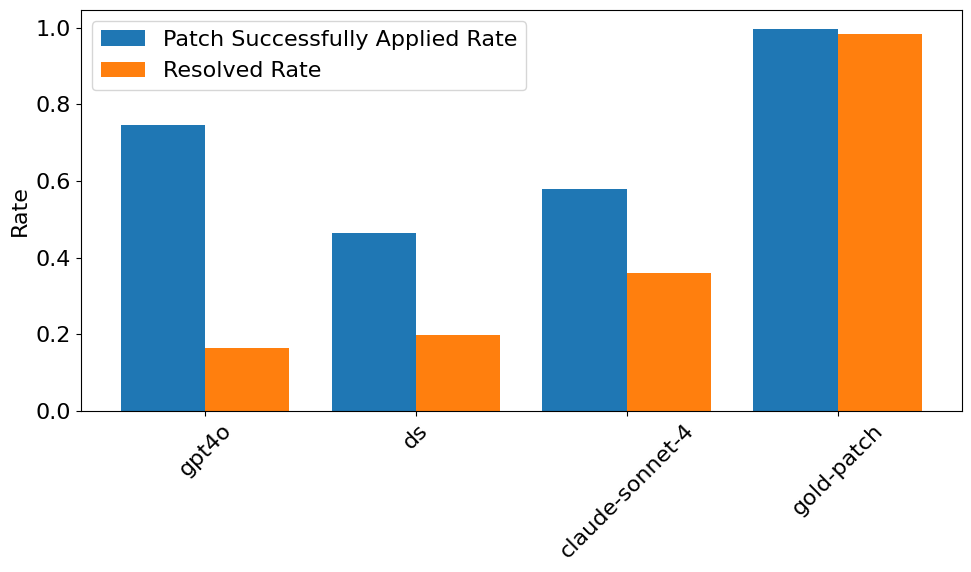}
    \caption{Patch generation results of different models in the baseline evaluation. The total number is 300.}
    \label{fig:1st_overview_barchart}
\end{figure}

\autoref{fig:1st_overview_barchart} presents the comparative performance of the three models and the gold patches across 300 input instances. The result shows clear differences among the three models across the two metrics: the \textbf{patch successfully applied rate (PSA)} and the \textbf{resolved rate}. The PSA indicates the proportion of generated patches that could be merged without errors, regardless of whether it compiles or passes the tests, and the resolved rate measures the proportion of patches that successfully resolved the issues described in the prompts. In all cases, their performance falls well below that of the gold patches, underscoring a gap in functional correctness between LLM-generated patches and the curated reference patches produced by humans. GPT-4o has the highest PSA rate (75\%) among the three models, but its resolved rate (16\%) is the lowest. Claude-Sonnet-4 achieves a PSA rate of 58\% together with the highest resolved rate of 36\% among the three models. DeepSeek shows a more moderate performance, with a PSA rate of 46\% and a resolved rate of 20\%. 

Among the failed attempts, we further inspected the agent logs to understand common failure causes. Notably, we observed that the agent's retries were largely devoted to resolving environment and shell command syntax errors (\eg import failures) instead of iteratively refining the patch logic, suggesting a weakness in context and tooling rather than code generation itself. 

\begin{figure*}[ht]
    \centering
    \includegraphics[width=1\linewidth]{ 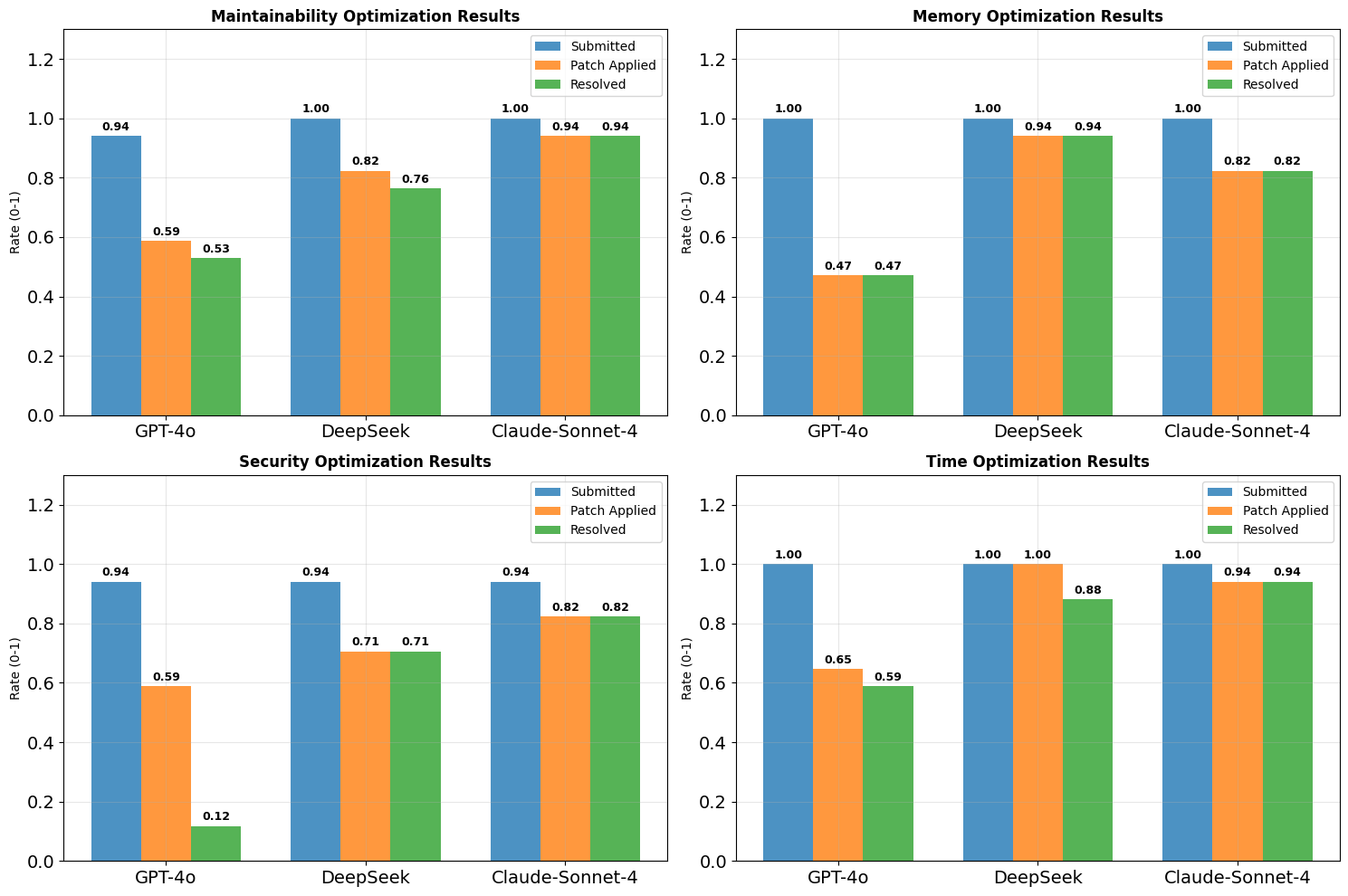}
    \caption{Patch generation results under different prompt strategies optimized for different NFQCs. Results are shown for three models across the selected metrics: (a) maintainability, (b) memory, (c) security, and (d) test runtime. For each metric, the bars illustrate the rate of submitted patches, the PSA, and the resolved rate. The total number is 17.}
    \label{fig:2nd_function_correctness}
\end{figure*}

\autoref{fig:2nd_function_correctness} reports functional correctness for patches regenerated with prompts tailored to specific NFQCs. Besides PSA and the resolved rate, we included the submitted rate, which denotes whether the model provides a prediction for an instance. Under these NFQC-specific prompts, the functional correctness of the previously resolved patches decreased. The security prompt induces the biggest degradation in correctness, while memory and time have a lower impact on functional correctness. For GPT-4o, the submitted rate drops from 100\% to 94\% under the maintainability prompt, indicating that the model failed to generate patches for some instances. Under the security prompt, all three models show reduced submitted rates, with GPT-4o having the lowest resolved rate (12\%). Claude-Sonnet-4 achieves the highest PSA and resolved rates under the maintainability, security, and time prompts, while DeepSeek-Reasoner achieves the best under the memory prompt, with both the PSA and the resolved rate reaching 94\%.

\paragraph{RQ 5.1: Baseline performance across the selected NFQCs}

To address RQ 5.1, we first evaluated the resolved generated patches from the first stage in terms of maintainability, security, runtime, and memory usage. To ensure comparability, the evaluation was conducted on the common subset of 17 resolved patches across the three models. \autoref{tab:codeql_rule_count} summarizes the CodeQL results on this common subset for maintainability and security, together with the average number of lines of code added by each patch. Following the official document of CodeQL, the rules are grouped by the severity level: \textbf{E}rror, \textbf{W}arning and \textbf{R}ecommendation.

Overall, the generated patches trigger substantially more CodeQL rules than the gold patches, with a substantially larger number of newly added issues reported for maintainability than for security. In terms of patch size, the gold patches introduce an average of 2.65 added lines of code, whereas the generated patches vary across models, ranging from 2.35 lines for DeepSeek, 5.94 for Claude-Sonnet-4, to 16.71 lines for GPT-4o.

For maintainability, the gold patches introduce 30 recommendations and no warnings or errors. In contrast, the generated patches trigger more than 400 maintainability recommendations per model (423 for GPT-4o, 420 for DeepSeek, and 414 for Claude-Sonnet-4), together with two warnings in each case. No maintainability errors are reported for either gold or generated patches. 

Additionally, the observed maintainability issues are not evenly distributed across rules. As shown in \autoref{tab:maintainability_rules}, the \textit{cyclic-import} rule alone accounts for 360 to 365 newly introduced issues in the patches generated by LLMs, while no such issues are observed in the gold patches. In contrast, the \textit{unused-global-variable} rule is introduced in comparable numbers by both gold and generated patches, with 30 occurrences in the gold patches and 35–36 occurrences across the generated patches. Other frequently triggered rules, such as \textit{ineffectual-statement} and \textit{unused-import}, occur only a small number of times and are absent from the gold patches. 

For security, the overall number of triggered rules remains low across all models. As reported in \autoref{tab:codeql_rule_count}, the gold patches trigger two security findings at the error level, while the generated patches trigger between four and six error-level findings. No security warnings or recommendations are reported for any patch type. All newly introduced security issues observed in the generated patches are attributable to the \textit{clear-text-storage-sensitive-data} rule.

\begin{table}[ht]
    \centering
        \caption{Distribution of maintainability issues across rules in gold and unoptimized generated patches.}
    \begin{tabular}{|p{3cm}|c|c|c|c|c|}
    \hline
       \textbf{RuleID}  & \textbf{Severity} &\textbf{Gold-patch }&\textbf{GPT-4o}&\textbf{DeepSeek}&\textbf{Claude} \\ \hline
       cyclic-import  &R &0&360&365&360 \\ \midrule
       unused-global-variable& R & 30&35&36&35 \\ \midrule
       ineffectual-statement & R &0&8&8&8 \\ \midrule
       unused-import & R &0 &5&5&5 \\ \midrule
       unused-local-variable &R &0&5&5&5\\ 

       \bottomrule
    \end{tabular}

    \label{tab:maintainability_rules}
\end{table}

\begin{table}[ht]
    \centering
        \caption{Results of CodeQL analysis on generated patches in terms of maintainability and security after the first stage. The table presents the number of rules triggered for each severity level across the generated patches and the gold patches. The severity level follows the order: \textbf{E}rror $>$ \textbf{W}arning $>$ \textbf{R}ecommendation.}

    \begin{tabular}{|c|c|c|c|c|c|c|c|}
        \hline
         & &\multicolumn{3}{c}{\textbf{Maintainability}} & \multicolumn{3}{|c|}{\textbf{Security}}  \\ \cline{3-8}
      \textbf{Patch} &\textbf{Avg Added LOC}& \textbf{E} & \textbf{W}&\textbf{R} & \textbf{E} & \textbf{W}&\textbf{R}  \\ \hline 
      \textbf{Gold-patch}&2.65 & 0 &0 &30  & 2 & 0&0  \\ \hline
      \textbf{GPT-4o}& 16.71& 0 &2 &423  & 4 & 0&0  \\ \hline 
      \textbf{DeepSeek}& 2.35& 0 &2& 420  & 6 & 0&0  \\ \hline
      \textbf{Claude-Sonnet-4}& 5.94& 0 & 2&414  & 4 & 0&0  \\ 

      \hline 
      
    \end{tabular}

    \label{tab:codeql_rule_count}
\end{table}

\autoref{tab:tm_stat} reports the performance of the unoptimized patches with respect to test runtime and memory usage. All statistics are computed over the common subset of 17 resolved patches. The generated patches from three models exhibit higher average test runtime and memory usage. GPT-4o records an average runtime of 23.37 seconds and a memory usage of 44.54 MB, both of which are clearly higher than the gold patches, which are 14.26 seconds and 23.50 MB. The patches generated by DeepSeek-Reasoner and Claude-Sonnet-4 show similar trends, with maximum test runtimes reaching 53.67 seconds and 51.21 seconds, respectively. 

\begin{table*}[ht]
    \centering
    \caption{Test runtime (in seconds) and memory usage (in MB) of unoptimized patches generated by three LLMs compared with the gold patches. For each model, the mean, standard deviation, minimum, and maximum values are reported across all 17 instances from the common subset.}

    \begin{tabular}{|c|c|c|c|c|c|c|c|c|}
    \hline 
        & \multicolumn{4}{c}{\textbf{Test Runtime} (s)} & \multicolumn{4}{|c|}{\textbf{Memory} (MB)}   \\ \hline

        \textbf{Model} & mean &std & min & max & mean &std & min & max \\ \hline 

        \textbf{Gold-patch} &14.26 & 12.29 & 3.51 &46.28 &23.50 & 3.84 &20.04 &37.25 \\ \hline 
        
        \textbf{GPT-4o}& 23.27 &19.75&5.64 & 68.77& 44.54&25.79&8.79&111.09 \\ \hline
        
       \textbf{DeepSeek}& 17.68 & 14.09&4.42&53.67 & 24.99 &7.22&9.25& 39.25  \\ \hline

        \textbf{Claude-Sonnet-4} & 16.72 &14.72&4.39&51.21& 27.27&9.91&8.20&48.61 \\ \hline

    \end{tabular}
    \label{tab:tm_stat}
\end{table*}

\paragraph{RQ 5.2: Effect of NFQC-specific Prompt}

To address RQ 5.2 and RQ 5.3, we applied NFQC-specific prompts to the patches generated in the first stage and re-evaluated the regenerated patches with respect to both functional correctness and the targeted NFQCs. 

On the common instance set, NFQC-specific optimization prompts are associated with reduced functional correctness. This effect is particularly evident for GPT-4o, where the resolved rate under the security optimization decreases to 0.12, corresponding to only two functionally correct patches.

As shown in \autoref{tab:gpt-instance}, for GPT-4o, within the common instance set, applying the corresponding NFQC-optimization prompts leads to different patterns across metrics. For maintainability, five out of nine instances exhibit fewer issues under the maintainability prompt,  while four instances show identical counts, and no instance shows an increase. For runtime, all 10 resolved instances show lower runtime under the runtime optimization prompt. For memory usage, seven out of eight instances exhibit lower memory consumption under the memory prompt, while one instance shows a higher value compared to the baseline. The security optimization only yielded two resolved instances, and we skipped the results.

\begin{longtable}{ccccccc}
\caption{Instance comparison for GPT-4o. M, S, T, and Me denote optimizations for Maintainability, Security, Runtime, and Memory, respectively. }\\
    \hline
    \toprule
       Instance id  & Metrics & Baseline & M & S& T & Me  \\
       \hline
       \endfirsthead
       
       \toprule
              Instance id  & Metrics & Baseline & M & S& T & Me  \\
    \midrule
       \endhead

       \midrule
       \multicolumn{7}{r}{\emph{Continue on next page}}
       \endfoot
       
       \bottomrule
       \endlastfoot

        \multirow{4}{*}{django-11039} & Maintainability& 10 & 10 & 10&14&2 \\
                                      & Security& 0 &0&0&0&0 \\
                                      & Time &38.81 &13.98&20.32&14.06&14.07 \\
                                      & Memory &42.87&21.34&21.32&222.50&21.35 \\ \midrule
        \multirow{4}{*}{django-11099} & Maintainability& 2 &- &- &0&0 \\
                                      & Security& 0 &-&-&0&0 \\
                                      & Time &23.36 &-&-&12.98&13.19 \\
                                      & Memory &77.76&-&-&174.80&20.62 \\ \midrule
        \multirow{4}{*}{django-11133} & Maintainability& 2 &- &- &-&- \\
                                      & Security& 0 &-&-&-&- \\
                                      & Time & 30.98&-&-&-&- \\
                                      & Memory &29.91&-&-&-&-\\   \midrule              
        \multirow{4}{*}{django-12700} & Maintainability& 2 & -& -&-&- \\
                                      & Security& 0 &-&-&-&- \\
                                      & Time & 51.15&-&-&-&- \\
                                      & Memory &51.15&-&-&-&- \\ \midrule
        \multirow{4}{*}{django-12983} & Maintainability& 2 & -& 2&2&2 \\
                                      & Security& 0 &-&0&0&0\\
                                      & Time &28.30 &-&12.07&13.15&9.48 \\
                                      & Memory &33.45&-&20.66&181.12&20.85 \\ \midrule
        \multirow{4}{*}{django-13658} & Maintainability& 2 &0&- &4&- \\
                                      & Security& 0 &0&-&0&- \\
                                      & Time &68.77 &46.44&-&49.60&- \\
                                      & Memory &32.58&22.62&-&182.20&- \\ \midrule
        \multirow{4}{*}{django-13710} & Maintainability& 2 & -&- &3&- \\
                                      & Security& 0 &-&-&0& -\\
                                      & Time &16.46 &-&-&8.23&- \\
                                      & Memory &111.09&-&-&195.51&- \\ \midrule
        \multirow{4}{*}{django-13933} & Maintainability& 2 & 2& -&2&2 \\
                                      & Security& 0 &0&-&0&0 \\
                                      & Time &5.64 &3.65&-&3.65&3.41 \\
                                      & Memory &20.57&18.81&-&177.27&18.84 \\ \midrule
        \multirow{4}{*}{django-14238} & Maintainability& 2 & 2& -&2& 2\\
                                      & Security& 0 &0&-&0&0 \\
                                      & Time &9.27 &6.59&-&6.34&6.16 \\
                                      & Memory &44.94&20.88&-&184.24&20.82 \\ \midrule
        \multirow{4}{*}{django-14382} & Maintainability& 2 & 0&- &2&0 \\
                                      & Security& 0 &0&-&0&0 \\
                                      & Time & 57.03&40.49&-&39.59&39.99 \\
                                      & Memory &31.91&23.78&-&168.48&21.41 \\ \midrule
        \multirow{4}{*}{django-15851} & Maintainability& 2 &0 &- &-&- \\
                                      & Security& 2 &0&-&-&- \\
                                      & Time &9.47 &6.20&-&-&- \\
                                      & Memory &33.84&145.75&-&-& -\\ \midrule
        \multirow{4}{*}{django-16046} & Maintainability& 2 & 0& -&2&0 \\
                                      & Security& 0 &0&-&2&0 \\
                                      & Time &7.61 &5.85&-&5.91&5.60 \\
                                      & Memory &26.41&20.14&-&179.16&20.13 \\ \midrule
        \multirow{4}{*}{django-16255} & Maintainability& 2 & 2&- &-&- \\
                                      & Security& 0 &0&-&-&- \\
                                      & Time &9.47 &5.89&-&-&- \\
                                      & Memory &28.59&20.74&-&-&- \\ \midrule
        \multirow{4}{*}{django-16527} & Maintainability& 3 & -& -&17& -\\
                                      & Security& 0 &-&-&0& -\\
                                      & Time &11.97 &-&-&6.64&- \\
                                      & Memory &85.62&-&-&189.80&- \\ \midrule
        \multirow{4}{*}{django-16595} & Maintainability& 2 & -&- &-&- \\
                                      & Security& 2 &-&-&-&- \\
                                      & Time &9.92 &-&-&-&- \\
                                      & Memory &57.53&-&-&-&- \\      \midrule
        \multirow{4}{*}{seaborn-3190} & Maintainability& 0 & -&- &-&- \\
                                      & Security& 0 &-&-&-&- \\
                                      & Time &5.71 &-&-&-&- \\
                                      & Memory &8.79&-&-&-&- \\  \midrule
        \multirow{4}{*}{xarray-5131	} & Maintainability& 386 &181 &- &-&181 \\
                                      & Security& 0 &0&-&-&0 \\
                                      & Time &11.66 &8.62&-&-&8.63 \\
                                      & Memory &40.15&16.95&-&-&109.20

    \label{tab:gpt-instance}
\end{longtable}

As shown in \autoref{tab:ds-instance}, for DeepSeek, within the common instance set, comparisons between the baseline and the NFQC-specific optimization show limited changes for maintainability and security, and mixed behavior for runtime and memory. For maintainability, one instance shows a reduction under the maintainability optimization, 12 instances remain unchanged, and one instance shows an increase. For security, one instance shows a reduction under the security optimization, 11 instances remain unchanged, and no instances show an increase. For runtime, 12 instances show lower runtime under the runtime optimization, while five instances exhibit higher runtime. Memory usage shows a split pattern. Eight instances exhibit lower memory usage under the memory optimization, while eight instances show higher memory usage.

\begin{longtable}{ccccccc}
\caption{Instance comparison for DeepSeek. M, S, T, and Me denote optimizations for Maintainability, Security, Runtime, and Memory, respectively.}\\
    \hline
    \toprule
       Instance id  & Metrics & Baseline & M & S& T & Me  \\
       \hline
       \endfirsthead
       
       \toprule
              Instance id  & Metrics & Baseline & M & S& T & Me  \\
    \midrule
       \endhead

       \midrule
       \multicolumn{7}{r}{\emph{Continue on next page}}
       \endfoot
       
       \bottomrule
       \endlastfoot
    
        \multirow{4}{*}{django-11039} & Maintainability& 2 &2 &2 &2&2 \\
                                      & Security& 0 &0&0&0&0 \\
                                      & Time &14.75 &20.66&17.50&20.95&20.98 \\
                                      & Memory &22.74&21.35&21.34&21.50&68.60 \\ \midrule
        \multirow{4}{*}{django-11099} & Maintainability& 2 &2 &8 &0&2 \\
                                      & Security& 0 &0&0&0& 0\\
                                      & Time &11.07 &19.80&16.44&20.01&20.13 \\
                                      & Memory &22.32&20.59&20.59&20.94&21.31 \\
                                      \midrule
        \multirow{4}{*}{django-11133} & Maintainability& 2 &2 &- &0&2 \\
                                      & Security& 0 &0&-&0& 0\\
                                      & Time & 10.20&19.43&-&19.65&19.80 \\
                                      & Memory &21.21&20.21&-&20.54&20.57 \\ \midrule               
        \multirow{4}{*}{django-12700} & Maintainability& 2 & 2&2 &2&3 \\
                                      & Security& 0 &0&0&0&0 \\
                                      & Time & 14.76&22.63&19.16&22.85&23.07 \\
                                      & Memory &22.90&21.71&21.70&26.43&30.38 \\ \midrule
        \multirow{4}{*}{django-12983} & Maintainability& 2 &3 &11 &2&- \\
                                      & Security& 0 &0&0&0&- \\
                                      & Time &10.01 &9.80&9.68&12.33& -\\
                                      & Memory &39.25&66.18&20.63&20.64&- \\ \midrule
        \multirow{4}{*}{django-13658} & Maintainability& 2 &- & 4&0&2 \\
                                      & Security& 0 &-&0&0&0 \\
                                      & Time &53.67 &-&47.11&49.30&46.30 \\
                                      & Memory &24.20&-&22.62&22.75&65.63 \\ \midrule
        \multirow{4}{*}{django-13710} & Maintainability& 2 &- &2 &2& 2\\
                                      & Security& 0 &-&0&0&0 \\
                                      & Time &22.61 &-&8.64&8.66&8.58 \\
                                      & Memory &33.37&-&21.19&21.19&30.49 \\ \midrule
        \multirow{4}{*}{django-13933} & Maintainability& 2 &2 &2 &2& 2\\
                                      & Security& 0 &0&0&0&0 \\
                                      & Time &4.42 &3.49&3.58&3.83&3.67 \\
                                      & Memory &28.74&18.82&18.82&18.96&20.26 \\ \midrule
        \multirow{4}{*}{django-14238} & Maintainability& 2 &2 &2 &2&2 \\
                                      & Security& 0 &0&0&0&0 \\
                                      & Time &14.39 &6.39&6.43&7.03&6.36 \\
                                      & Memory &36.42&20.80&20.80&136.30&24.67 \\ \midrule
        \multirow{4}{*}{django-14382} & Maintainability& 2 &2 & 2&0& 2\\
                                      & Security& 0 &0&0&0& 0\\
                                      & Time & 52.11&40.06&41.59&40.25&40.68 \\
                                      & Memory &22.54&21.42&21.41&21.59&21.95 \\ \midrule
        \multirow{4}{*}{django-15851} & Maintainability& 2 &2 &- &0&2 \\
                                      & Security& 2 &7&-&0&0 \\
                                      & Time &12.05 &6.18&-&6.15&6.17 \\
                                      & Memory &21.51&20.42&-&20.52&20.67 \\ \midrule
        \multirow{4}{*}{django-16046} & Maintainability& 2 & 2&2 &0&2 \\
                                      & Security& 2 &2&2&2&2 \\
                                      & Time &10.91 &5.75&5.95&5.77&6.00 \\
                                      & Memory &21.23&20.14&20.14&20.16&26.15 \\ \midrule
        \multirow{4}{*}{django-16255} & Maintainability& 2 &2 &- &2&2 \\
                                      & Security& 2 &2&-&2&2 \\
                                      & Time &13.00 &5.92&-&6.06&5.91 \\
                                      & Memory &21.98&20.75&-&20.75& 25.10\\ \midrule
        \multirow{4}{*}{django-16527} & Maintainability& 2 &2 & 2&-&2 \\
                                      & Security& 0 &0&0&-&0 \\
                                      & Time &10.33 &6.52&6.85&-&6.66 \\
                                      & Memory &33.24&23.45&23.42&-& 37.63\\ \midrule
        \multirow{4}{*}{django-16595} & Maintainability& 2 &- & 3&0&2 \\
                                      & Security& 2 &-&1&0&0 \\
                                      & Time &18.38 &-&5.86&5.96&5.59 \\
                                      & Memory &24.17&-&22.12&35.02&22.66 \\ \midrule     
        \multirow{4}{*}{seaborn-3190} & Maintainability& 0 &- &- &-&0 \\
                                      & Security& 0 &-&-&-&0 \\
                                      & Time &5.93 &-&-&-&4.49 \\
                                      & Memory &9.25&-&-&-&62.25 \\  \midrule
        \multirow{4}{*}{xarray-5131	} & Maintainability& 392 &183 & -&183&183 \\
                                      & Security& 0 &0&-&0&0 \\
                                      & Time &21.92 &8.56&-&8.53&9.35 \\
                                      & Memory &19.78&16.98&-&17.02&84.03

    \label{tab:ds-instance}
\end{longtable}

\autoref{tab:sonnet-instance} shows the results of Claude-Sonnet-4. Within the common instance set, the application of NFQC-optimization prompts leads to limited changes for maintainability and security, and reductions for runtime and memory in most instances. For maintainability, one instance shows a reduction under the maintainability optimization, 13 instances remain unchanged, and two instances show higher counts. For security, one instance shows a reduction, and one instance shows an increase under the security optimization, while 12 instances remain unchanged. The median security count remains zero for both baseline and security optimization. For runtime, 14 instances show lower runtime under the runtime optimization, while two instances exhibit higher runtime. 
For memory usage, 14 instances exhibit lower values under the memory optimization, and no instances show higher memory usage compared to the baseline.

\begin{longtable}{ccccccc}
\caption{Instance comparison for Claude-Sonnet-4. M, S, T, and Me denote optimizations for Maintainability, Security, Runtime, and Memory, respectively.}\\
    \hline
    \toprule
       Instance id  & Metrics & Baseline & M & S& T & Me  \\
       \hline
       \endfirsthead
       
       \toprule
              Instance id  & Metrics & Baseline & M & S& T & Me  \\
    \midrule
       \endhead

       \midrule
       \multicolumn{7}{r}{\emph{Continue on next page}}
       \endfoot
       
       \bottomrule
       \endlastfoot

        \multirow{4}{*}{django-11039} & Maintainability& 2 &2 &2 &2&2 \\
                                      & Security& 0 &0&0&0&0 \\
                                      & Time &12.74 &17.34&13.93&17.27&14.09 \\
                                      & Memory &24.40&21.34&21.36&21.35&21.34 \\ \midrule
        \multirow{4}{*}{django-11099} & Maintainability& 2 & -&2 &2&2 \\
                                      & Security& 0 &-&0&0&0 \\
                                      & Time &30.42 &-&13.01&16.43&13.15 \\
                                      & Memory &29.11&-&20.58&20.59&20.59 \\ \midrule
        \multirow{4}{*}{django-11133} & Maintainability& 2 &6 &- &-&- \\
                                      & Security& 0 &0&-&-&- \\
                                      & Time & 34.10&15.95&-&-&- \\
                                      & Memory &22.81&20.21&-&-& -\\   \midrule             
        \multirow{4}{*}{django-12700} & Maintainability& 2 &2 &- &2&- \\
                                      & Security& 0 &4&-&0&- \\
                                      & Time & 26.88&18.81&-&19.12&- \\
                                      & Memory &24.31&21.71&-&21.71& -\\ \midrule
        \multirow{4}{*}{django-12983} & Maintainability& 2 &2 &2 &2&2 \\
                                      & Security& 0 &0&0&0&0 \\
                                      & Time &9.14 &9.49&13.12&9.45& 13.05\\
                                      & Memory &24.57&20.64&20.64&20.66&20.64 \\ \midrule
        \multirow{4}{*}{django-13658} & Maintainability& 2 &2 &2 &2&2 \\
                                      & Security& 0 &0&0&0&0 \\
                                      & Time &51.21 &46.22&49.63&46.14&49.79 \\
                                      & Memory &25.50&22.63&22.65&22.64& 22.88\\ \midrule
        \multirow{4}{*}{django-13710} & Maintainability& 2 &2 &2 &2&2 \\
                                      & Security& 0 &0&0&0&0 \\
                                      & Time &12.23 &8.12&8.09&8.21&8.01 \\
                                      & Memory &31.60&21.20&21.20&21.20& 21.32\\ \midrule
        \multirow{4}{*}{django-13933} & Maintainability& 2 & 2& 2&2&2 \\
                                      & Security& 0 &0&0&0&0 \\
                                      & Time &4.39 &3.43&3.52&3.49&3.54 \\
                                      & Memory &20.27&18.82&18.82&18.82&18.82 \\ \midrule
        \multirow{4}{*}{django-14238} & Maintainability& 2 & 2& 2&2&- \\
                                      & Security& 0 &0&0&0& -\\
                                      & Time &7.69 &6.24&6.25&6.23&- \\
                                      & Memory &33.86&20.83&20.80&21.02&- \\ \midrule
        \multirow{4}{*}{django-14382} & Maintainability& 2 &2 &2 &2&10 \\
                                      & Security& 0 &0&0&0&0 \\
                                      & Time & 43.74&40.03&39.94&39.91&40.46 \\
                                      & Memory &26.14&21.61&21.43&21.43&21.44 \\ \midrule
        \multirow{4}{*}{django-15851} & Maintainability& 2 &2 &2 &2&2 \\
                                      & Security& 0 &0&0&2&2 \\
                                      & Time &6.92 &6.08&5.91&6.03&5.98 \\
                                      & Memory &24.88&20.41&20.41&20.43&20.42 \\ \midrule
        \multirow{4}{*}{django-16046} & Maintainability& 2 &2 &13 &5&2 \\
                                      & Security& 0 &0&0&0&0 \\
                                      & Time &6.95 &5.80&5.66&5.81&5.78 \\
                                      & Memory &30.76&20.13&20.13&20.15&20.14 \\ \midrule
        \multirow{4}{*}{django-16255} & Maintainability& 2 &2 &9 &2&2 \\
                                      & Security& 0 &2&2&0&2 \\
                                      & Time &7.05 &5.88&5.99&5.93&6.05 \\
                                      & Memory &22.27&20.80&20.74&20.75& 20.75\\ \midrule
        \multirow{4}{*}{django-16527} & Maintainability& 3 &44 &2 &2&2 \\
                                      & Security& 0 &2&0&0&2 \\
                                      & Time &9.35 &6.49&6.39&6.43&6.55 \\
                                      & Memory &48.61&23.45&23.42&23.42& 23.44\\ \midrule
        \multirow{4}{*}{django-16595} & Maintainability& 2 &2 &2 &2&2 \\
                                      & Security& 2 &2&0&0&0 \\
                                      & Time &8.10 &5.53&5.53&5.49&5.56 \\
                                      & Memory &48.58&145.16&22.12&22.12&22.12 \\  \midrule    
        \multirow{4}{*}{seaborn-3190} & Maintainability& 0 &0 & -&0&0 \\
                                      & Security& 0 &0&-&0&0 \\
                                      & Time &4.54 &4.68&-&4.10&4.05 \\
                                      & Memory &8.20&146.84&-&8.03&8.03 \\  \midrule
        \multirow{4}{*}{xarray-5131	} & Maintainability& 386 &181&181 &181&181 \\
                                      & Security& 0 &0&0&0&0 \\
                                      & Time &8.71 &8.39&8.42&8.45&8.27 \\
                                      & Memory &17.69&17.06&17.02&17.09& 16.95 
    \label{tab:sonnet-instance}
\end{longtable}

\begin{figure*}[ht]

    \centering
  \begin{subfigure}[t]{0.5\textwidth}
    \centering
    \includegraphics[width=\linewidth]{ 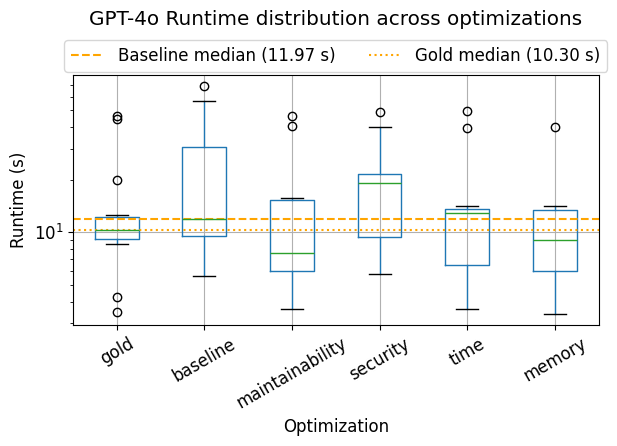}
    \label{fig:gpt_bp_t}
  \end{subfigure}\hfill
  \begin{subfigure}[t]{0.5\textwidth}
    \centering
    \includegraphics[width=\linewidth]{ 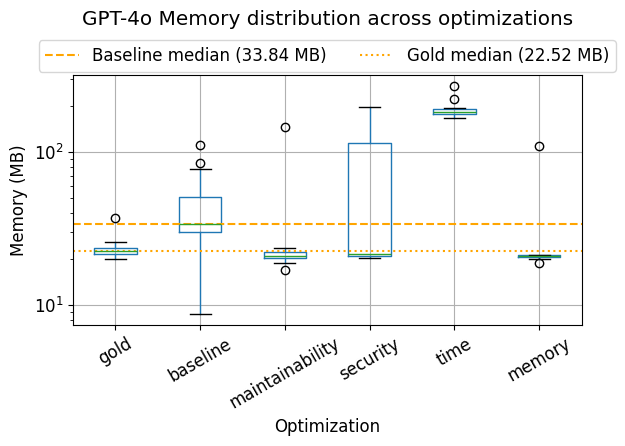}
    \label{fig:gpt_bp_m}
  \end{subfigure}
    \begin{subfigure}[t]{0.5\textwidth}
    \centering
    \includegraphics[width=\linewidth]{ 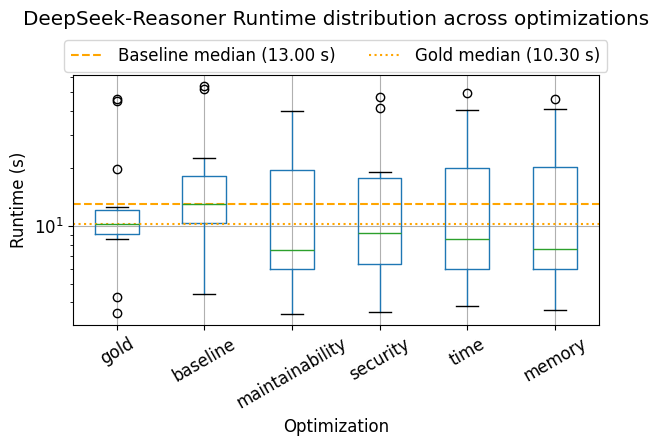}
    \label{fig:ds_bp_t}
  \end{subfigure}\hfill
  \begin{subfigure}[t]{0.5\textwidth}
    \centering
    \includegraphics[width=\linewidth]{ 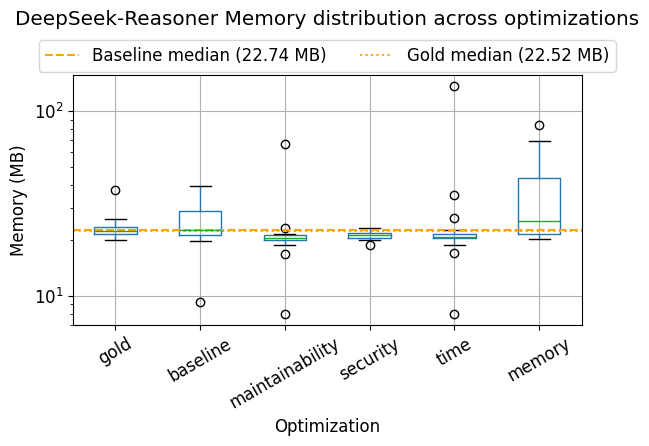}
    \label{fig:ds_bp_m}
  \end{subfigure}
    \begin{subfigure}[t]{0.5\textwidth}
    \centering
    \includegraphics[width=\linewidth]{ 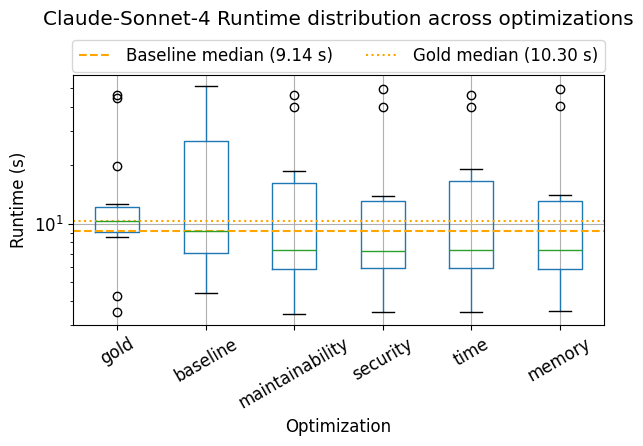}
    \label{fig:sonnet_bp_t}
  \end{subfigure}\hfill
  \begin{subfigure}[t]{0.5\textwidth}
    \centering
    \includegraphics[width=\linewidth]{ 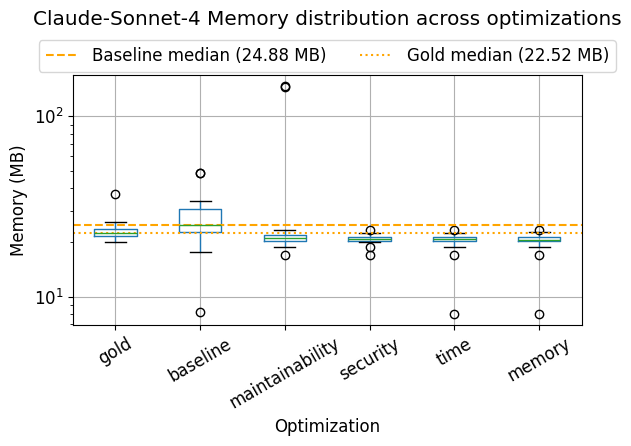}
    \label{fig:sonnet_bp_m}
  \end{subfigure}
  
  \caption{Boxplots of test runtime (s) and memory usage (MB) on a logarithmic scale for patches generated by GPT-4o, DeepSeek-Reasoner, and Claude-Sonnet-4, compared with the gold and baseline patches. The \textbf{baseline} represents the results of the unoptimized generated patches of each model. The boxes summarize the distribution across resolved instances, with dashed and dotted lines indicating the median values of the baseline evaluation and gold patches, respectively.}
  \label{fig:bxp}
\end{figure*}

\autoref{fig:bxp} shows the distribution of test runtime and memory usage across different optimization prompts for three models, with gold and baseline patches included as reference points. Across all three models, the runtime distributions show noticeable variation across optimization prompts and individual instances. For GPT-4o and DeepSeek-Reasoner, median test runtimes differ across optimization settings, with some prompts associated with lower medians than the baseline and others showing higher medians. However, the IQRs (Interquartile Range) overlap with those of the gold and baseline patches in all cases, indicating that runtime differences are not uniform across instances. For Claude-Sonnet-4, median test runtimes across optimization prompts remain close to the gold reference, while the baseline setting exhibits greater variability. Outliers are observed under most settings for all models, highlighting the presence of instances with notably higher test runtimes.

Memory usage distributions similarly exhibit overlap across gold, baseline, and optimization-specific prompts for all models. GPT-4o shows larger variability in memory usage, particularly under the baseline and security prompts, whereas the maintainability and memory prompts yield median values closer to the gold patches. For DeepSeek-Reasoner, memory optimization results in increased dispersion, while maintainability and time optimization yield some outliers with higher memory usage. For Claude-Sonnet-4, memory usage remains relatively stable across all optimization settings, with medians closely aligned with the gold patches and limited dispersion, aside from a small number of outliers.

 We further analyzed instances where the generated patches under the time optimization setting achieve lower test runtime than the gold patches (10.30 s and 22.52 MB), including \textit{django-16046} and \textit{django-16527}. For these instances, the generated patches introduce simpler and more direct changes that reduce runtime while increasing memory usage. The gold patches, in contrast, make use of existing methods or functions of the project, which results in higher runtime but lower memory usage. This difference in implementation helps explain why the generated patches outperform the gold patches in test runtime for these cases.

\paragraph{RQ 5.3: Trade-offs across NFQCs}

According to the results reported in \autoref{tab:gpt-instance}, \autoref{tab:ds-instance}, and \autoref{tab:sonnet-instance}, for some instances, changes in one NFQC coincide with changes in others, indicating interactions and trade-offs between different NFQCs.

For GPT-4o, although maintainability optimization does not reduce maintainability issues for all generated patches, the patches produced under maintainability optimization exhibit lower runtime and memory usage than the baseline patches for most instances, with the exception of \textit{django-15851}. Runtime optimization tends to degrade the other three NFQCs, especially for memory usage. For example, in the case of \textit{django-13933}, runtime optimization results in a rise from 20.57 MB to 177.27 MB in memory usage compared to the baseline generated patches. Compared with runtime optimization, memory optimization improves the runtime and memory usage of eight resolved instances, but increases the memory usage of \textit{xarray-5131}.

For DeepSeek-Reasoner, maintainability and security optimizations lead to increases in maintainability and security issue counts in a small number of instances. Notably, the \textit{xarray-5131} shows improvements in maintainability under the three NFQC optimizations except security. Runtime optimization produces more mixed effects. It improves maintainability, security, and memory usage for the majority of instances, but a few cases (\eg \textit{django-14238}) exhibit clear regressions.

Claude-Sonnet-4 exhibits a similar pattern. Maintainability optimization leads to increased maintainability issue counts in a small number of instances (django-16527 and django-11133), and it also results in a substantial increase in memory usage for \textit{django-16595}.

\begin{table}[htbp]
\centering
\caption{Significance Test Results and Effect Sizes ($r$) across Models. We report optimization strategies that achieve both statistical significance ($p < 0.05$) and large effect sizes ($r > 0.5$). For each metric, we additionally report the minimum $p$-value and the maximum effect size observed across optimization strategies. $-$ indicates that no optimization met the reporting criteria for the corresponding metric. The best results are highlighted in bold.}
\label{tab:significance_summary}
\begin{tabular}{llccc}
\toprule
\textbf{Model} & \textbf{Metric} & \textbf{Optimization} & \textbf{$p$-value} & \textbf{Effect($r$)} \\ \midrule
\multirow{4}{*}{GPT-4o} & Maintainability & - & 0.0625 & 0.659 \\
 & Time & M / T / Me & \textbf{0.0020} & \textbf{0.979}   \\
 & Memory & T & \textbf{0.0020} & \textbf{0.979}   \\
 & Security & - & 1.0000 & 0.000 \\ \midrule
\multirow{4}{*}{DeepSeek} & Maintainability & T & \textbf{0.0039} & \textbf{0.700}  \\
 & Time & - & 0.0640 & 0.535 \\
 & Memory & S & \textbf{0.0005} & \textbf{1.000}  \\
 & Security & - & 0.1573 & 0.354  \\ \midrule
\multirow{4}{*}{Sonnet}  & Maintainability & - & 0.5930 & 0.143  \\
 & Time & T & \textbf{0.0042} & \textbf{0.716}   \\
 & Memory &S / T / Me  &  \textbf{0.0001} &  \textbf{1.000}  \\
 & Security & - & 0.1025 & 0.408  \\ \bottomrule
\end{tabular}
\end{table}

 \autoref{tab:significance_summary} shows the results of our statistical analysis using the Wilcoxon signed-rank test. The results show statistically significant and practically meaningful improvements primarily for runtime and memory metrics. 

For GPT-4o, runtime improvements are statistically significant across Maintainability, Time and Memory optimization settings ($p = 0.0020, r = 0.979$), and time optimization yields a significant increase in memory usage ($p = 0.0010, r = 0.99$). No maintainability optimization reaches statistical significance, although large effect sizes are observed for some settings. No statistically significant effects are observed for security-related metrics.

For DeepSeek, statistically significant improvements are observed for maintainability under the time optimization ($p = 0.0039, r = 0.70$) and for memory under the security optimization ($p = 0.0005, r = 1.00$). Runtime improvements show a large effect size but do not reach conventional significance thresholds ($p = 0.0640, r = 0.54$), and no statistically significant effects are detected for security-related metrics.

For Claude-Sonnet-4, statistically significant improvements are observed for runtime under the time optimization ($p = 0.0042, r = 0.72$) and for memory under the security, time, and memory optimizations ($p = 0.0001, r = 1.00$). In contrast, maintainability and security optimizations do not yield statistically significant differences.

\subsection{Discussion of experiment results}

\begin{figure*}[ht]
    \centering
    \includegraphics[width=0.9\textwidth]{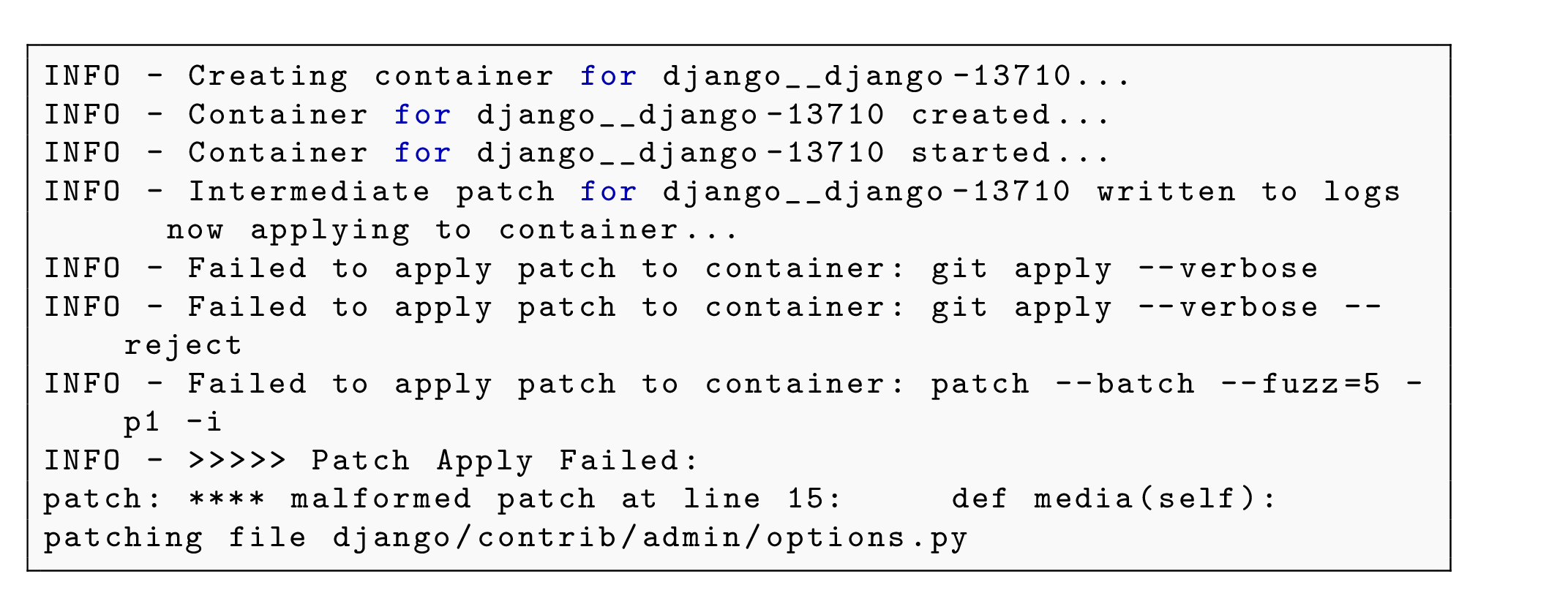}

    \caption{Example of malformed patch error during patch application.}
    \label{fig:malformed_log}
\end{figure*}

The results demonstrate that current code LLMs face substantial limitations when addressing complex engineering tasks. By simulating how developers realistically use LLMs in their workflows, our evaluation provides insights into their actual applicability in software development. As discussed in \autoref{experiment_results}, many of the agent retries were spent on resolving environment and command issues rather than the patch logic. These issues reveal a weakness in environment configuration, dependency resolution, and command invocation. In an agentic and robust system, the ability to use tools effectively is an integral part of the overall capability; therefore, such failures indicate limitations at the process level, even when the code generation itself is adequate. Moreover, in the final stage, many patches failed to apply due to context mismatches (see \autoref{tab:mismatch} and \autoref{fig:malformed_log}), highlighting persistent difficulties in adhering to the unified diff format despite clear instructions in the prompts.

\begin{table}[t]
    \centering
    \caption{Number of malformed patches generated by each model under different optimization prompts (out of 17 patches per optimization). A malformed patch refers to a generated patch that could not be successfully applied to the repository or failed to comply with the unified diff format.}
    
    {\tabcolsep=1.6mm
    \begin{tabular}{|c|c|c|c|c|}
    \hline
      \textbf{Model / Optimization}   & \textbf{Maintainability}& \textbf{Security} & \textbf{Time} & \textbf{Memory} \\ \hline
      GPT-4o &3&4&3& 7\\ \hline 
      DeepSeek-Reasoner &2&0&0&0 \\ \hline 
      Claude-Sonnet 4 &0&1&1&3 \\ \hline
    \end{tabular}
    }
    \label{tab:mismatch}
\end{table}

The comparison across models reveals different patterns. GPT-4o achieved the highest PSA rate in the first evaluation, yet its resolved rate was the lowest, suggesting that many of its patches were syntactically valid but functionally ineffective. After introducing NFQC optimizations, both its application and resolution rates declined, showing limited robustness in balancing functional correctness and NFQCs. In contrast, Claude-Sonnet-4 exhibited a lower PSA in the first evaluation but achieved a higher resolved rate. After the NFQC optimization, Claude-Sonnet-4 achieved the highest PSA and resolved rates under maintainability, security, and runtime optimization, while DeepSeek-Reasoner performed best under memory optimization, with both PSA and resolved rate reaching 94\%. These results suggest a specialization of strengths among models across different optimization objectives, rather than a single model outperforming others universally.

A further challenge lies in the lack of NFQC awareness. Under baseline prompts, all models produced patches that were inferior to the gold patches in maintainability, security, runtime, and memory usage. This suggests that agent-based generation pipelines, while capable of producing functionally correct patches, implicitly prioritize ``passing the test'' rather than ``passing with quality.'' Such a lack of NFQC awareness constitutes a fundamental barrier to the integration of LLMs into automated workflows. It is not simply that models are unaware of best practices; their decision mechanisms do not treat NFQCs as primary objectives. Since LLMs are trained to predict the most likely next token, they capture superficial patterns of code rather than underlying engineering principles. Without an intrinsic cost function for code quality, all test-passing solutions are treated as equivalent. As a result, directly applying these patches risks introducing technical debt: the code may be functionally correct but structurally fragile and suboptimal. This limitation reflects the current model architectures and training paradigms rather than implementation errors.

Our baseline analysis showed that LLM-generated patches triggered more maintainability rules than the gold patches. For example, in \autoref{tab:codeql_rule_count}, GPT-4o triggers 423 maintainability recommendations compared to only 30 in the gold patch. Results from DeepSeek-Reasoner and Claude-Sonnet-4 reinforced this pattern. Additionally, the maintainability issues in gold patches are \textit{unused-global-variable}, but most of the maintainability issues in generated code are \textit{cyclic-import}. The gold patches are typically written with awareness of the existing project structure and therefore tend to reuse established modules and dependency patterns. As a result, their maintainability issues are often localized, such as \textit{unused-global-variable}. In contrast, generated patches often resolve defects in a more self-contained manner, introducing or reorganizing imports without fully considering the global dependency graph. In large Python projects, this can easily lead to \textit{cyclic-import} issues, which substantially increase the number of maintainability warnings reported by CodeQL. This observation also highlights a broader challenge faced by current LLM-based software development. Even when models are allowed to access and explore the entire code repository through prompting, they may still produce errors that stem from an incomplete understanding of the project as a whole. Such errors suggest that current models lack robust project-level awareness, and future research should place greater emphasis on supporting holistic reasoning over large and evolving codebases.

The NFQC optimization results also revealed trade-offs between different NFQCs. We observed interactions between runtime and memory usage in some instances, where improvement in runtime is at the cost of memory usage. More critically, under the optimization prompts used in this study, we observed negative optimization, where models degraded the very dimension they were instructed to improve. These results indicate that optimizing a single NFQC through prompts does not guarantee improvements in isolation. Instead, changes introduced in order to optimize one dimension may interact with other NFQCs or, in some cases, negatively affect the targeted metric itself. This suggests that current NFQC optimization primarily operates through local code transformations that are sensitive to the surrounding code context, rather than through a stable understanding of cross-metric constraints. As a result, NFQC optimization cannot be treated as a reliable, standalone procedure. Any optimization attempt should be accompanied by targeted validation, as improvements observed for one NFQC or one instance cannot be assumed to generalize across metrics or across instances.

Finally, two particular instances stand out in our results. \textit{xarray-5131} showed consistent improvements in maintainability and security across models and optimizations, suggesting that instances with severe initial deficiencies and relatively lenient functional constraints allow for modifications without breaking tests. By contrast, \textit{seaborn-3190} illustrated the inherent conflict among NFQCs. Across several optimizations, functional correctness was compromised, suggesting that its code was in a fragile equilibrium where any modification risked undermining both security and functionality.

\section{Threats to Validity}
\label{sec:threats}

This section presents the potential issues that may affect the validity and reproducibility of our findings.

\subsection{Internal Validity}

For the literature review, our search strategy followed key elements of the PRISMA guidelines \cite{Page2020TheP2}, but there is still a risk that relevant studies have been missed due to database coverage limitations or the keywords. To mitigate the risk, we applied both forward and backward snowballing and found several relevant studies.

For the workshops, two authors independently took notes during the workshops; individual interpretations may have introduced subjective bias and affected the objectivity of the notes. To mitigate this, we conducted joint discussions after each workshop to reconcile the two records, ensuring a consistent interpretation of the practitioners' views.

In our experiments, we combined the prompts provided in the SWE-bench Lite benchmark with the CodeQL reports to design NFQC optimization prompts. This procedure mitigates subjective bias to some extent, but prompt sensitivity remains a threat. Minor changes in the prompts could affect LLM behavior and result in different generated patches. In addition, different hyperparameter settings, such as temperature, may influence the stochasticity and diversity of model outputs.

Second, our experiments were conducted on a single workstation (AMD Ryzen 7, 16GB RAM, Ubuntu 22.04). Although the evaluation environment was kept consistent across all models and stages, background system loads could have introduced small variations in runtime and memory measurements. These potential fluctuations were not explicitly recorded. 

Third, we used three LLMs in the experiment. But the models update frequently during the experiment period, which could lead to model drift and different behaviors in the future. This is a known limitation when evaluating LLMs.

Finally, to ensure comparability across models,  we restricted our analysis to instances successfully solved by three models. However, it may create a bias towards simpler tasks, potentially skewing the observed trade-offs among NFQCs.

\subsection{Construct Validity}
For the literature review, we mapped diverse quality characteristics from different studies to the ISO/IEC 25010 quality model; there is a risk that we may not always understand and reflect the authors' original intent during the mapping. 

Individual understanding of concepts such as readability and security by participants might also have varied, leading to potential inconsistencies in the way in which some discussions were understood. To mitigate this, we reviewed and discussed our notes to ensure a consistent understanding of their perspectives.

The metrics used in the experiment represent practical but partial approximations of NFQCs. In particular, we rely on CodeQL findings to operationalize maintainability and security of generated patches. As a static analysis tool, CodeQL provides reproducible indicators of certain code-level quality issues, but it captures only a subset of the underlying quality dimensions. To partially mitigate this limitation, we additionally report patch size and conduct a selective qualitative human review of representative generated patches, which provides complementary perspectives on code changes and helps contextualize the automated results. Nevertheless, we acknowledge that CodeQL-based measurements cannot fully reflect broader aspects of code quality, such as readability, architectural coherence, or developer-perceived maintainability. Future work should therefore incorporate more comprehensive evaluation approaches, including additional static and dynamic analysis tools as well as more extensive human-centered assessment, to obtain a more holistic view of non-functional code quality.

\subsection{External Validity}

Most of the workshop participants are Software Center companies, and there is a risk that their perspectives may not fully represent the broader software industry.

The generalizability of our findings is constrained by both the dataset and the experimental design. First, our study is based on SWE-bench Lite, which consists of real-world issue-resolution tasks. This benchmark is well-suited for studying NFQCs in repository-level patches, but the observed patterns may not extend to other generation settings, such as feature implementation or class-level synthesis \cite{Du24evaluating,Li25FEA}. At the same time, compared to synthetic benchmarks (e.g., HumanEval), SWE-bench Lite requires multi-file navigation, dependency analysis, and interaction with existing codebases, which more closely reflect professional software maintenance scenarios. Additionally, the instances included in SWE-bench are skewed toward framework-based Python repositories, particularly Django (127 Django out of 300 instances). This framework dominance influences both the nature of the vulnerabilities and the manifestation of NFQCs, as many security features are abstracted away by the framework, and maintainability challenges differ from those in lower-level codebases.

Second, the analysis is restricted to instances that were successfully resolved by all evaluated models. This restriction biases the sample toward tasks that are jointly solvable and therefore likely easier than the full benchmark. As a result, the findings should be interpreted as characterizing non-functional quality behavior under relatively favorable conditions. Since quality degradation and trade-offs are more likely to emerge in complex or borderline cases, the reported results may underestimate the extent of NFQC issues that arise in harder tasks.

Third, the final dataset contains 17 instances. While limited in size, this selection enables detailed qualitative inspection, including manual review and patch-size analysis, which would be difficult to apply consistently at larger scales. The study, therefore, emphasizes analytical depth over coverage.

Finally, although the specific models evaluated reflect a snapshot of rapidly evolving code generation systems, the observed patterns, such as tensions between functional success and structural maintainability, are likely indicative of broader tendencies in current LLM-based code generation approaches. Extending the analysis to larger and more diverse task sets, including instances not jointly solved by all models, remains an important direction for future work.

\section{Conclusion and Future Work}
\label{conclusion}
\subsection{Conclusion}

Our study examined the performance of LLM-generated code with respect to NFQCs in real-world software engineering tasks. By combining evidence from a literature review, industry workshops, and empirical experiments, we provide a multi-perspective view of how NFQCs are currently conceptualized, prioritized, and manifested in LLM-assisted code generation.

The literature review shows that existing research has mainly focused on security, performance efficiency, maintainability, and reliability, while other NFQCs have received comparatively little attention. At the same time, different studies often employ inconsistent terminology for the same NFQC, which hampers cross-study comparisons and highlights the absence of a unified framework. 

Insights from workshops highlight the differences between academic and industrial perspectives. While academic research tends to emphasize security and performance efficiency, practitioners show stronger concern for maintainability, particularly in the context of large software systems and long-term maintenance. 

Our experiments explore the interactions between NFQCs in generated code, demonstrating the instability of optimizing NFQCs through single-prompt code generation in practical software engineering settings, and showing that applying LLM-generated patches to projects without validation entails the risk of introducing and amplifying technical debt.

Taken together, these findings illustrate existing limitations and trade-offs in how NFQCs are handled in current LLM-based code generation workflows. Rather than drawing general conclusions about LLM capabilities in all settings, this study highlights the importance of aligning evaluation frameworks, optimization strategies, and validation practices with the quality concerns that arise in real-world software development.

\subsection{Future Work}

To move the generated code from ``passing the tests'' to ``passing with quality'', future work should connect evaluation, modeling, and tooling. In future work, we are going to integrate strong verification mechanisms into the generation pipeline. We will embed static analysis and automated detectors into generation and optimization so that NFQCs are monitored in real time, used as gates for candidate patches, and returned as structured feedback for revision. By addressing NFQCs before deployment, we aim to reduce technical debt accumulation and improve the reliability of LLM-generated code in software systems.

\section{Acknowledgment}

This work was funded by the Software Center project 61 and the Vinnova Competence Center for Continuous Digitalization 2023-00546. The authors are indebted to Willem Meijer for fruitful discussions.

\bibliography{reference,SecondFinal}
\bibliographystyle{elsarticle-harv}

\end{document}